\documentclass [preprint,
 amsmath,amssymb,
 aps,
 superscriptaddress,
12pt,floatfix]{revtex4-2}
\usepackage[title]{appendix}
\usepackage[T1]{fontenc}
\usepackage{lmodern}

\usepackage{graphicx}
\usepackage{capt-of}
\usepackage{geometry}
\usepackage[monochrome]{xcolor}
\usepackage{amsmath}
\usepackage{placeins}
\usepackage{array}

\usepackage{tabularray}
\usepackage{makecell}
\usepackage{setspace}

\usepackage{url}
\usepackage{breakurl}

\usepackage{xr}

\usepackage{algorithm}
\usepackage{algpseudocode}

\begin{document}

\title{A model-based assessment of social isolation practices for COVID-19 outbreak response in residential care facilities}

\author{Cameron Zachreson*}
\affiliation{School of Computing and Information Systems, The University of Melbourne, Parkville, Victoria, Australia}
\email{cameron.zachreson@unimelb.edu.au}

\author{Ruarai Tobin}
\affiliation{Centre for Epidemiology and Biostatistics, Melbourne School of Population and Global Health, The University of Melbourne, Parkville, Victoria, Australia}

\author{Camelia Walker}
\affiliation{School of Mathematics and Statistics, The University of Melbourne, Parkville, Victoria, Australia}

\author{Eamon Conway}
\affiliation{The Walter and Eliza Hall Institute, Melbourne, Australia}

\author{Freya M Shearer}
\affiliation{Centre for Epidemiology and Biostatistics, Melbourne School of Population and Global Health, The University of Melbourne, Parkville, Victoria, Australia}

\author{Jodie McVernon}
\affiliation{Victorian Infectious Disease Reference Laboratory Epidemiology Unit, The Royal Melbourne Hospital at the Peter Doherty Institute for Infection and Immunity, Melbourne, VIC, Australia}
\affiliation{Department of Infectious Diseases, The University of Melbourne at the Peter Doherty Institute for Infection and Immunity, Melbourne, VIC, Australia}

\author{Nicholas Geard}
\affiliation{School of Computing and Information Systems, The University of Melbourne, Parkville, Victoria, Australia}

\date{\today}

\begin{abstract} 

{\bf{Background: }}{Residential aged-care facilities (RACFs, also called long-term care facilities, aged care homes, or nursing homes) have elevated risks of respiratory infection outbreaks and associated disease burden. During the COVID-19 pandemic, social isolation policies were commonly used in these facilities to prevent and mitigate outbreaks. We refer specifically to {\color{purple} general isolation} policies that were intended to reduce contact between residents, without regard to confirmed infection status. Such policies are controversial because of their association with adverse mental and physical health indicators and there is a lack of modelling that assesses their effectiveness. }\\
{\bf{Methods: }}{In consultation with the Australian Government Department of Health and Aged Care, we developed an agent-based model of COVID-19 transmission in a structured population, intended to represent the salient characteristics of a residential care environment. Using our model, we generated stochastic ensembles of simulated outbreaks and compared {\color{purple} summary statistics of outbreaks simulated} under different mitigation conditions. {\color{purple} Our study focuses on the marginal impact of general isolation (reducing social contact between residents), regardless of confirmed infection.} For a realistic assessment, our model included other generic interventions consistent with the Australian Government's recommendations released during the COVID-19 pandemic: isolation of confirmed resident cases, furlough {\color{purple}(mandatory paid leave)} of staff members with confirmed infection, and deployment of personal protective equipment (PPE) after outbreak declaration.}\\
{\bf{Results: }}{{\color{purple} In the absence of any asymptomatic screening, general isolation of residents to their rooms reduced median cumulative cases by approximately 27\%. However, when conducted concurrently with asymptomatic screening and isolation of confirmed cases, general isolation reduced the median number of cumulative infections by only 12\% in our simulations.} }\\
{\bf{Conclusions: }}{{\color{purple}Under realistic sets of assumptions, our simulations showed that} general isolation of residents did not provide substantial benefits beyond those achieved through screening, isolation of confirmed cases, and deployment of PPE. Our results also highlight the importance of effective case isolation, and indicate that asymptomatic screening of residents and staff may be warranted, especially if importation risk from the outside community is high. {\color{purple}Our conclusions are sensitive to assumptions about the proportion of total contacts in a facility accounted for by casual interactions between residents.} }\\

\end{abstract}

\maketitle

\section{Keywords}

residential aged care; nursing homes; long-term care; agent-based model; COVID-19; outbreak response; social isolation; nonpharmaceutical intervention; infectious disease dynamics; contact network

\section{Background}

Residential aged-care facilities (RACFs, also called long-term care facilities, aged care homes, or nursing homes) have elevated risks of respiratory infection outbreaks and associated disease burden. Heightened risk arises due to importation of pathogens by visitors and staff, close and prolonged contact among residents {\color{purple}(occupants of the facility)} and staff members, and the demographic and health profiles of residents. These intersecting factors produce scenarios in which outbreaks are difficult to mitigate and carry disproportionate consequences in terms of medical impact. Simultaneously, the physical infrastructure, insecure workforce, high health needs of residents, and particular social requirements of RACFs impose limitations on the types of control measures that can be enacted without compromising the mental health of residents \cite{usher2021preparedness}. {\color{purple} In contrast with other types of healthcare settings, RACFs feature long-term stays (multiple years \cite{zhang2023expected}), frequent social interactions between residents, and a large part-time workforce with variable levels of medical training \cite{RACFcensus}.}

Modelling efforts made early during the COVID-19 pandemic did not emphasize the confluence of risk factors present in aged-care environments {\color{purple}(e.g., high prevalence of comorbidity and limited infection control capacity)} that would later be acknowledged as responsible for large numbers of preventable deaths \cite{pagel2022role}. This led to a substantial and ongoing global research effort applying infectious disease modelling to outbreak detection, response, and prevention in residential care scenarios \cite{smith2020optimizing,delaunay2020evaluation,wilmink2020real,vilches2021multifaceted,fosdick2022model,nguyen2022hybrid,lasser2021agent}. 

While modelling improved substantially following the acknowledgement of the disproportionate clinical significance of COVID-19 outbreaks in aged care, the strength of evidence for intervention effectiveness from observational and modelling studies has still been assessed as generally weak \cite{stratil2021non,sims2022social}. This has so far limited the quality of information available to policy makers faced with challenging and complex decisions about COVID-19 response measures in aged care. 

In particular, there is a lack of modelling work that specifically assesses the effectiveness of general isolation conditions applied to resident populations, despite these being possibly the most controversial policies implemented across the sector globally. {\color{purple} By ``general isolation''} we refer specifically to policies that were intended to reduce contact between residents, without regard to confirmed infection status. {\color{purple}These include, for example, shutting down shared meal facilities and closing social spaces at facilities, or encouraging residents to stay in their rooms}. Around the world, though such policies were not consistently included in government recommendations, many research articles, clinical reports, interviews, media reports, and informal testimony agree that facilities implemented general isolation measures in attempts to reduce the impact of ongoing outbreaks \cite{rios2020preventing,danilovich2020nursing,dyer2022managing,sweeney2022experiences,liddell2021isolating,Haj2020high}.

Here we apply an agent-based model of outbreak mitigation in aged care, to investigate the effectiveness of such policies, and assess the conditions under which they may be justified. For a realistic assessment, our model includes other generic interventions consistent with the Australian Government's recommendations released during the COVID-19 pandemic: isolation of confirmed resident cases, furlough {\color{purple}(mandatory paid leave)}  of staff with confirmed infection, and deployment of personal protective equipment (PPE) after outbreak declaration \cite{outbreak_guidelines}.


\section{Methods}
\subsection{overview}
In consultation with the Australian Government Department of Health and Aged Care, we developed an agent-based model of COVID-19 transmission in a structured population, building upon and modifying earlier modelling frameworks used to assess international arrival and quarantine pathways \cite{ZachresonQuar2022}. {\color{purple}In general, structured populations feature spatial and temporal clustering in contact patterns between individuals determined by, for example, heterogeneous but persistent dwelling locations and activity schedules. For such populations, homogeneous assumptions about contact patterns are not appropriate for detailed simulations of infectious disease transmission dynamics. To capture these important structural properties,} key features of the model include detailed representation of:
\begin{itemize}
    \item Facility characteristics, including resident and staff numbers;
    \item Contact patterns among staff and residents, incorporating details of staff scheduling;
    \item Infection dynamics, including time-varying infectiousness and test sensitivity;
    \item Screening and response strategies.
\end{itemize}

To capture the limitations of early detection, we implemented realistic screening and response strategies. These were based on those set out in the Australian Government Department of Health document ``COVID-19 Outbreaks in Residential Care Facilities, Communicable Disease Network Australia, National Guidelines for the Prevention, Control and Public Health Management of COVID-19 Outbreaks in Residential Care Facilities'' dated February 15th, 2022 \cite{outbreak_guidelines}. Full details of the model implementation and assumptions are provided in the {\color{blue}Supporting Information}.  

\subsection{facility structure}
\begin{figure}
    \centering
    \includegraphics[width=0.6\textwidth]{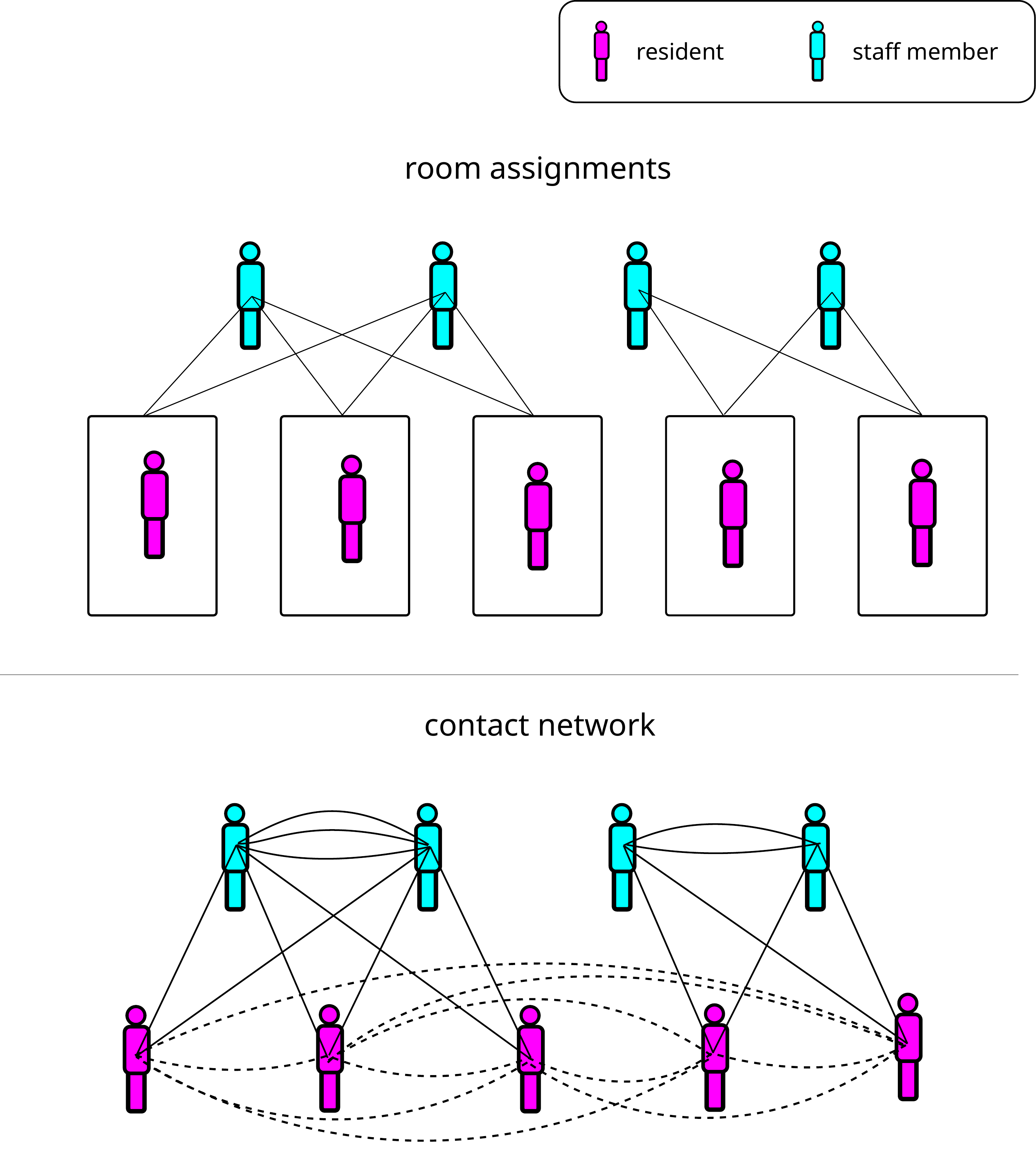}
    \caption{To generate a structured contact network, residents are assigned to rooms in which they live and staff members are assigned to sets of rooms that they service. This is done to ensure that each room is serviced by the same number of staff. Potential contacts are then created between any two individuals who are assigned to the same room. This creates a network structure from which contacts are sampled during simulation of transmission dynamics. In the contact network depicted here, solid lines represent edges derived from room assignments while dashed lines represent potential casual contacts between residents.}
    \label{fig:network_schematic}
\end{figure}

Briefly, a facility is modelled as a static population of residents, allocated to single-occupancy rooms. For this study, we simulated a facility with 121 staff and 88 residents, which is a typical size for an Australian RACF \cite{FacilitySize}. Staff attend a facility according to a weekly roster, with each staff member working five, three, or two days per week which accounts for a total of 73.8 full-time equivalent (FTE) staff (in Australia, 1.0 FTE corresponds to 5 full-time shifts in one week). Staff members are allocated to a set of rooms housing residents they will visit during the days they attend a facility. These allocations are used to generate a network of potential contacts, from which potentially infectious contact events are randomly sampled (Figure~\ref{fig:network_schematic}). In addition to these structured contacts, random contacts between residents in different rooms are included separately to allow for interactions mediated by communal areas and group activities that are not explicitly simulated in the facility model. Infectious contacts are subject to a probabilistic transmission process that depends on the time since the transmitting case was infected and any mitigation measures in place (see {\color{blue}Supporting Information}).

\subsection{within-host model of infection}
Infection proceeds through an incubation period, followed by symptom expression and eventual recovery (see {\color{blue} the Supporting Information} for more details on the model of individual infectiousness trajectories). Each infected individual is potentially infectious {\color{blue} from the moment of exposure, through} the incubation period and the duration of symptoms. Infectiousness increases until approximately the time of symptom onset, and then declines until recovery. We assume that, {\color{blue} on average,} 33\% of infections never express symptoms {\color{blue}and that symptom expression does not alter infectiousness given contact with someone who is susceptible. That is, the time-dependent trajectory of infectiousness is identical for symptomatic and asymptomatic infections. However, case detection following symptom onset may alter contact patterns due to infection control procedures, see below.} {\color{purple} The assumed asymptomatic fraction of 33\% is consistent with global estimates, however, we do not account for heterogeneity by age which could slightly increase the likelihood of case detection in residents relative to staff members. \cite{shang2022percentage,ma2021global,oran2021proportion,sah2021asymptomatic}.}

\subsection{baseline transmission dynamics}
\begin{figure}[h]
    \centering
    \includegraphics{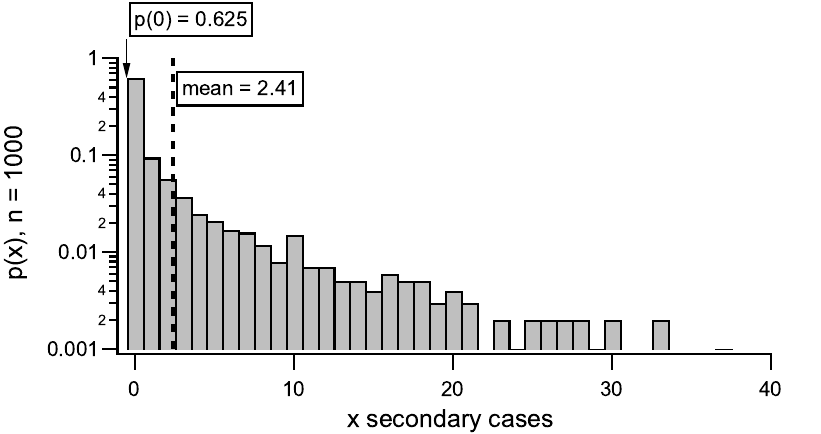}
    \caption{Index case secondary case distribution for 1000 realisations. }
    \label{fig:overdisp}
\end{figure}

In line with observed SARS-CoV-2 transmission dynamics, the model includes overdispersion of case infectiousness, in which approximately 60\% of index cases produce no secondary cases, while approximately 10\% of index cases produce more than ten secondary cases (Figure~\ref{fig:overdisp}) \cite{tariq2020real,endo2020estimating}. 

We calibrated the disease transmission model to produce a reproductive ratio of $R_0 \approx 2.4$, to simulate a scenario consistent with SARS-CoV-2 transmission in populations with modest immune-derived protection. These conditions correspond approximately to early estimates of the effective reproduction number of the Omicron variant \cite{liu2022effective}. Further, the reproductive ratio we selected is broadly consistent with the assumptions used in previous modelling studies  \cite{vilches2021multifaceted,love2021continued,smith2020optimizing}. {\color{blue} See the Supporting Information for more details on calibration.}

\subsection{initialisation}
Each simulation commences with a single index case and runs until there are no infections remaining in a facility or, in the case that an outbreak has been declared, the outbreak is declared to be over. Index cases are selected at random from the combined population of staff and residents. We note that this initialisation means that we are not simulating continuous importation of cases from the community. This choice excludes from the scope of our study any assessment on the role of visitation restrictions, community prevalence, or the behaviour of staff members outside of the facility. Rather, our model focuses on the transmission dynamics produced by single index cases.

\subsection{outbreak detection and response}
\begin{figure}[h]
    \centering
    \includegraphics[width = 0.6\textwidth]{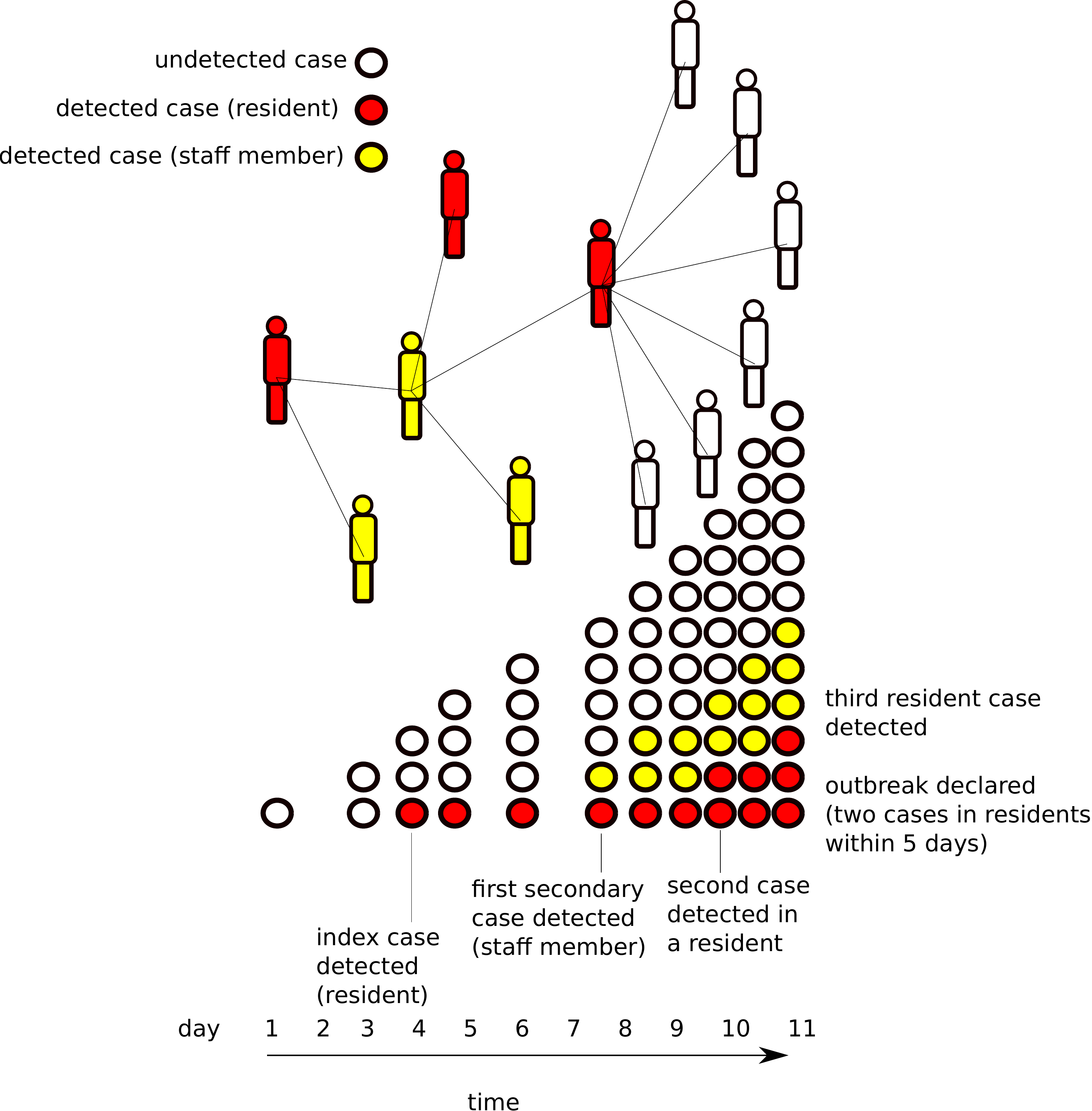}
    \caption{Schematic of the transmission process and outbreak declaration after detection of a specified number of cases within a given time window. Outbreaks are declared either from the detection of two cases in residents within five days, or detection of five cases in staff members within seven days. }
    \label{fig:declaration}
\end{figure}

\begin{figure}[h]
    \centering
    \includegraphics[width = 0.8\textwidth]{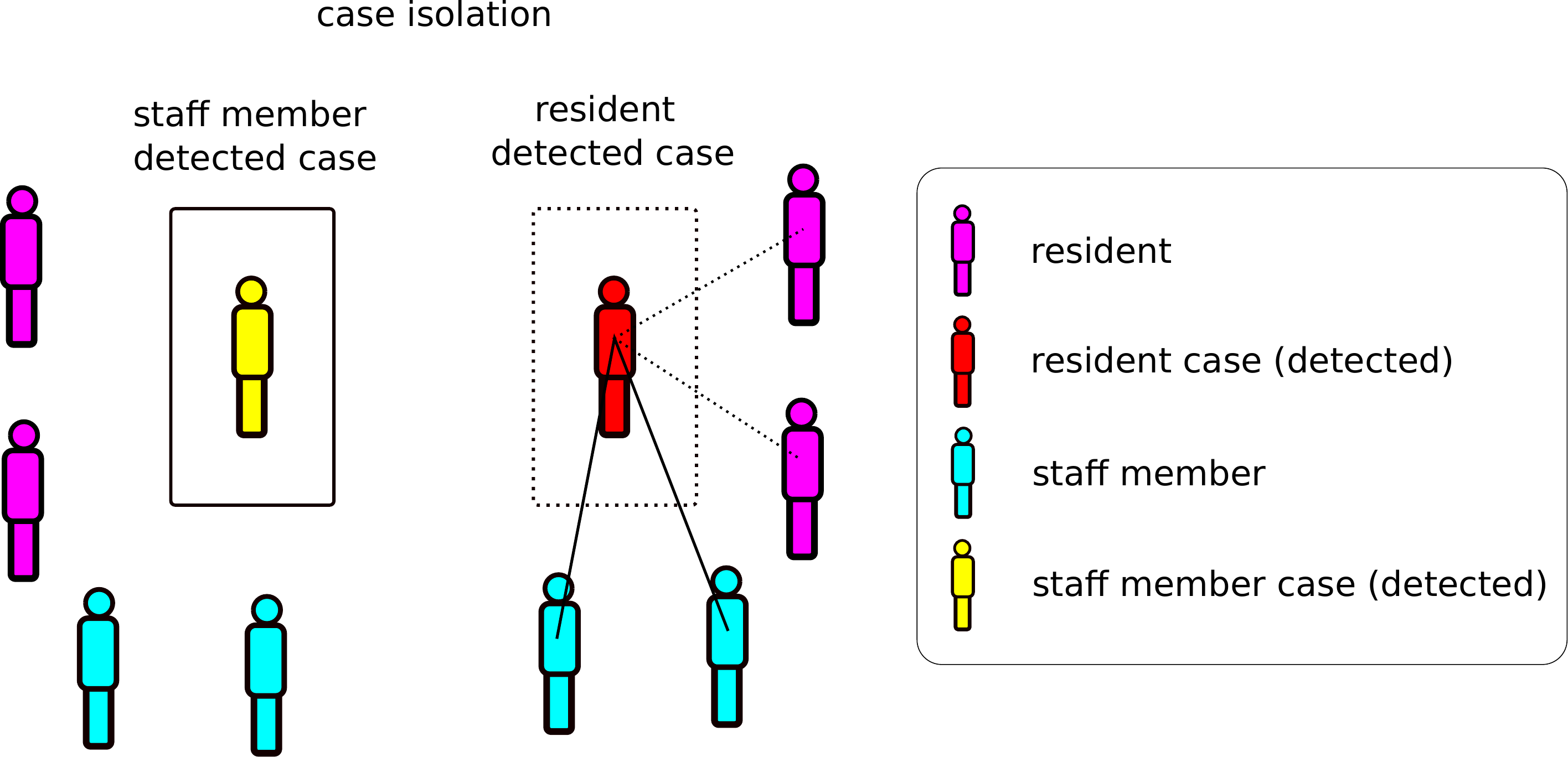}
    \caption{Schematic of case isolation procedure after detection through a positive test result. Residents remain in contact with staff members at the same frequency regardless of case isolation, but have reduced levels of contact with other residents (contact frequency with residents in different rooms reduced by 90\%). Isolated staff members are completely removed from contact (furloughed, not present at the facility). }
    \label{fig:isolation}
\end{figure}

\begin{figure}[h]
    \centering
    \includegraphics[width = \textwidth]{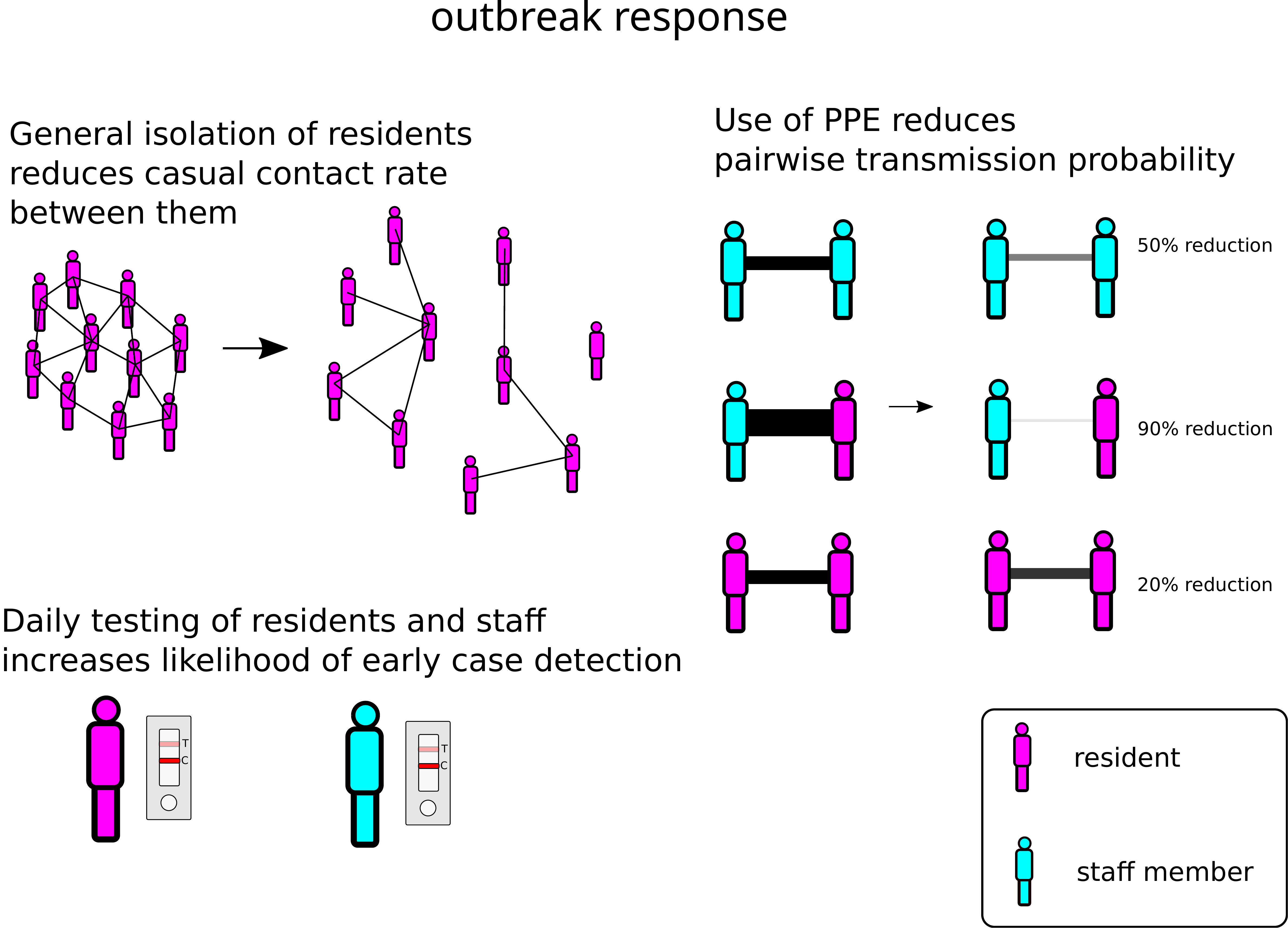}
    \caption{After an outbreak is declared, several different types of responses may follow. General isolation of residents reduces the rate of random contacts between them by a fixed proportion (which we vary in this study). Asymptomatic testing schedules are adjusted after the outbreak is declared, depending on the screening strategy simulated. Additionally, the deployment of PPE reduces the force of infection (pairwise transmission rate), depending on the types of individuals coming into contact, with a 90\% reduction between staff and residents, a 50\% reduction between staff members, and a 20\% reduction between residents. }
    \label{fig:schematic_outbreak_response}
\end{figure}

The model includes asymptomatic testing of staff and residents, as well as testing on development of symptoms. Infected residents are (imperfectly) isolated from other residents upon detection, but still have contact with staff. Infected staff are furloughed for seven days upon detection, and have no further contact within a facility during the furlough period (Figure~\ref{fig:isolation}). As per the Australian Government's guidelines (dated 15th February, 2022 \cite{outbreak_guidelines}), an outbreak is declared once two cases have been detected in residents within a five day period or five cases have been detected among staff members over a seven day period (Figure~\ref{fig:declaration}). Once an outbreak has been declared, the frequency of testing of staff and residents may increase and infection control measures (e.g., use of PPE) are put in place, reducing the probabilities of transmission among staff and residents (Figure~\ref{fig:schematic_outbreak_response}). {\color{purple} We assume differential effects of these infection control measures depending on the type of contact. For contacts between residents, we assume a relatively limited effectiveness of 20\% due to factors such as lower compliance with physical distancing, inconsistent mask wearing, and limited training in the proper use of PPE. For contacts among staff members, we assume a medium effectiveness of 50\% due to higher levels of compliance and training with PPE, but inconsistent infection control during contact events not involving residents. For contacts between staff members and residents, we assume the highest level of effectiveness (90\%) due to high levels of compliance and training with PPE protocols (including the proper use of N95 masks) for staff members when interacting with residents. These assumptions are broadly in line with literature estimates of PPE efficacy (see, e.g., \cite{schoberer2022rapid}). See the Supporting Information for more details related to model implementation of outbreak response measures.}  

\subsection{simulated policies}
 We simulated the following set of policies for outbreak detection through asymptomatic screening: 
\begin{enumerate}
\item{Full asymptomatic screening: prior to outbreak declaration, staff are tested once or twice per week, for part-time and full-time employees, respectively. Residents are tested every day (noting that daily screening of residents is not a recommendation of the Australian Government's guidelines, this represents an extreme screening scenario). During outbreaks, daily testing continues for residents, and staff are tested every day they are present as determined by the facility roster.}
\item{Asymptomatic screening during outbreaks: staff and residents are not subject to asymptomatic screening unless an outbreak is declared, after which they are subject to daily screening (as above).}
\item{No asymptomatic screening: testing is only conducted after symptom expression. No asymptomatic tests are conducted, regardless of outbreak status.}
\item{Unmitigated outbreaks: no testing is conducted. This means outbreaks are never declared, cases are never detected and isolated, and the transmission dynamics proceed without any mitigation measures. This `worst case' scenario provides a baseline comparison against which intervention effectiveness can be assessed.}
\end{enumerate}

For each of the above screening policies, we vary the extent to which the resident population (regardless of case status) is socially isolated from one another following declaration of an active outbreak. Here, we present results for 0\% (no general isolation), 50\% (partial isolation), and 90\% (stringent isolation) reductions in background contact rate. Note that residents in case isolation always have their background contact rate reduced by 90\% (all scenarios). {\color{purple} The assumption that case isolation is imperfect arises from the limitations to isolation reported in residential aged care settings due to the high prevalence of cognitive impairment and impulsive mobility behaviour, as well as the practical and legal limits to non-consensual physical restraint (see e.g., \cite{gilbert2020covid,liddell2021isolating,iaboni2020achieving}).}

This produces a set of 10 scenarios. For each screening strategy (other than the unmitigated outbreak scenario) we perform simulations until 1000 outbreaks have been simulated (the total number of simulations required for this can vary depending on the screening strategy). This produces an outbreak size distribution that is unimodal because it does not include those simulations in which transmission dies out early on {\color{blue}and no outbreak is declared}. For the unmitigated scenario (in which outbreaks are not declared), a fixed number of independent simulations are performed (10,000 instances). This produces a bimodal distribution of outbreak sizes because it includes those simulations in which transmission dies out early (Figure \ref{fig:I_tot_CIPPE_vs_UNMIT}a). We report summary statistics for the second mode (large outbreaks) to facilitate comparison between unmitigated and declared outbreaks. For each scenario, we report the following summary statistics: 

\begin{itemize}
\item {cumulative infections: the total number of infections (whether or not they are detected)}
\item {outbreak duration: the amount of time between outbreak declaration and termination.}
\item {peak staffing deficit: the maximum total full-time-equivalent simultaneously missing from the facility due to furlough of staff who test positive.}
\item {cumulative residents isolated: the total number of residents placed into case isolation due to a positive test result.}
\item {time to outbreak declaration: the time between introduction of the index case and declaration of the outbreak.}
\item {time to first detection: the time between introduction of the index case and the first case detection.}
\end{itemize}

The total number of infections is a generic measure of outbreak severity, but does not account for time-dependent properties of an outbreak. We report the outbreak duration as a measure of severity because costly and burdensome mitigation measures are in place throughout this period. The total number of residents placed into case isolation due to confirmed infection provides an additional measure of severity relevant both to apparent infection numbers and also to detrimental impacts of social isolation. The time to outbreak declaration and the time to first detection do not vary as functions of outbreak response, and provide estimates of the benefits of asymptomatic screening for early detection. The peak staffing deficit (measured in FTE) is a measure of the labour shortage (absenteeism) in the workforce due to furlough of staff. We chose to report the peak FTE deficit rather than the cumulative deficit because it provides a more intuitive measure of the logistical stress experienced within the facility due to personnel shortages. For example, if one staff member working five shifts in one week were furloughed, this would produce 1.0 FTE deficit. If one staff member working two days per week were furloughed, this would produce a 2/5 FTE deficit.

\FloatBarrier
\section{Results}

\subsection{The effects of asymptomatic screening strategies}

\begin{figure}[h]
    \centering
    \includegraphics[width = \textwidth]{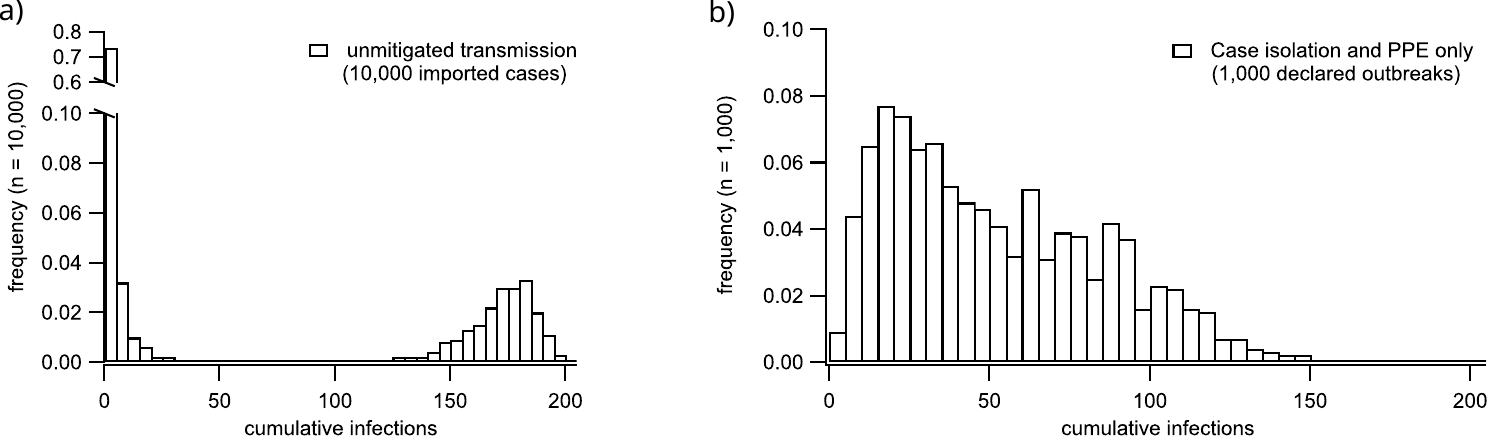}
    \caption{Cumulative infection distributions for simulated RACF outbreaks that are unmitigated (a), and which are subject to isolation of detected cases and deployment of PPE (b). The distribution of case numbers in (a) is produced over 10,000 independent stochastic simulations and includes those which are subject to early stochastic die-out. In these unmitigated scenarios, no screening takes place, no infections are detected, and no mitigation protocols are implemented. The histogram in (b) includes only those introductions which trigger the declaration of an outbreak, and show the case distribution over 1000 such outbreaks. More final size statistics are provided in the supporting material.}
    \label{fig:I_tot_CIPPE_vs_UNMIT}
\end{figure}

With a reproduction ratio of $R_0 \approx 2.4$, unmitigated outbreaks can infect large proportions of the facility population. As is typical of stochastic SIR-type epidemic dynamics, outbreak sizes follow a bimodal distribution separated between those that undergo stochastic die-out during their early stages, and those that progress to large-scale epidemics. A histogram of the final size distribution for unmitigated outbreaks is provided in Figure \ref{fig:I_tot_CIPPE_vs_UNMIT}(a). In contrast, even without asymptomatic testing or general isolation of residents, the use of PPE and case isolation produces a substantial drop in the distribution of outbreak size. A histogram of cumulative infection incidence for outbreaks with only PPE and case isolation in place is shown in Figure \ref{fig:I_tot_CIPPE_vs_UNMIT}(b).

\begin{figure}[h]
    \centering
    \includegraphics[width = \textwidth]{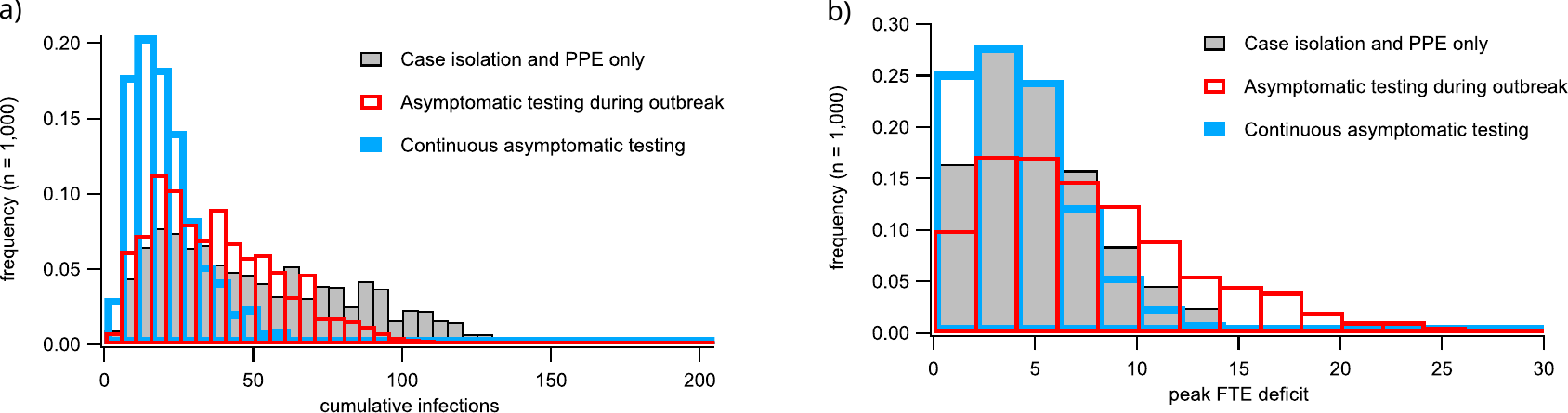}
    \caption{Histograms of cumulative infections (a) and peak FTE deficit due to furlough of staff members with confirmed infection (b). Grey bars correspond to scenarios employing isolation of symptomatic cases confirmed through testing and deployment of PPE after outbreak declaration. Red bars correspond to scenarios in which the declaration of outbreaks triggers asymptomatic screening of staff and residents. Blue bars correspond to scenarios in which continuous asymptomatic screening is conducted prior to outbreak declaration.}
    \label{fig:CIPPE_vs_OB_only_vs_Asymp}
\end{figure}

Introducing asymptomatic screening after the outbreak is declared reduces infection incidence by detecting cases earlier, but increases the peak FTE deficit (Figure \ref{fig:CIPPE_vs_OB_only_vs_Asymp}). This is because staff who are pre-symptomatic or asymptomatic upon the initiation of the outbreak are detected within a short time window after asymptomatic screening is initiated, resulting in more staff simultaneously furloughed. In the most extreme screening scenario, with asymptomatic testing beginning before the outbreak is detected, infections are mitigated while also avoiding the increase in FTE deficit associated with outbreak response. However, such strategies require diligent screening regimes, with large numbers of tests conducted even in the absence of any detected cases.

\subsection{The effects of general isolation conditions}
While effective at mitigating outbreaks, continuous asymptomatic screening of residents and staff may not be feasible under normal circumstances. The complete absence of asymptomatic screening or its implementation only during an active outbreak are more plausible scenarios, so we use them to illustrate the effects of introducing general isolation of residents. Here, we investigate to what extent the severity of outbreaks can be mitigated by general isolation conditions in combination with screening, case isolation, and PPE deployment. 

\begin{figure}[h]
    \centering
    \includegraphics[width = \textwidth]{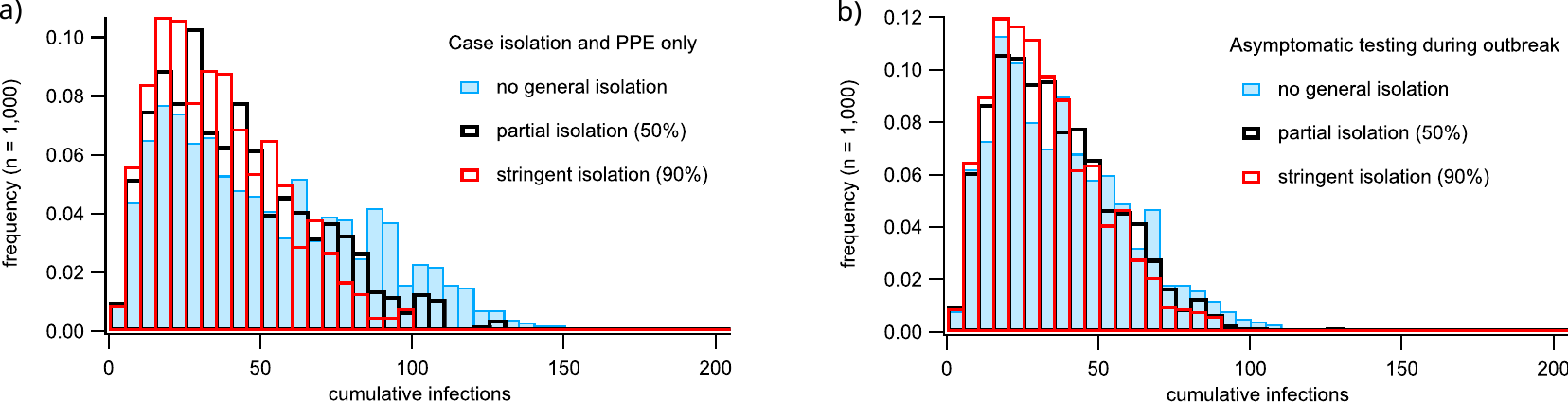}
    \caption{Histograms of cumulative infection numbers for outbreak scenarios involving varying levels of enforcement of a general resident isolation policy. Subfigure (a) shows statistics for scenarios in which outbreak response includes only isolation of symptomatic infections confirmed through testing and deployment of PPE after outbreak declaration. Subfigure (b) shows statistics for scenarios including asymptomatic screening after outbreak declaration. The different colored bars correspond to varying levels of isolation stringency, implemented as reductions in the rate of casual contact between residents. }
    \label{fig:CIPPE_vs_OB_only_vs_LD}
\end{figure}

Our results demonstrate that, while partial and full isolation of the resident population can make a marginal difference in the absence of asymptomatic screening (Figure \ref{fig:CIPPE_vs_OB_only_vs_LD}a), it has a negligible effect when combined with asymptomatic screening during outbreaks (Figure \ref{fig:CIPPE_vs_OB_only_vs_LD}b). This somewhat surprising finding is due in part to the effectiveness of asymptomatic screening, targeted case isolation, and PPE use, and in part to a fundamental constraint to isolation of aged care residents: that residents must remain in contact with facility staff during outbreaks. Therefore, even if residents are prevented from interacting with one-another, their interactions with staff members can result in transmission. This, combined with the relatively substantial reductions in transmission from the use of PPE and the isolation of confirmed cases, leaves little to be achieved through general isolation of the resident population. For comparison with Figure \ref{fig:CIPPE_vs_OB_only_vs_LD}, frequency distributions of cumulative cases for scenarios in which all outbreak mitigation procedures (PPE, case isolation, and asymptomatic testing) are disabled are shown in the Supporting Information Figure~S3. This comparison demonstrates that reducing contact between residents can mitigate against transmission and makes a substantial difference to infection totals when other response measures are not implemented. 

{\color{purple} Because we model general isolation as a reduction in the contact rate between residents, the marginal impact of general isolation is sensitive to the rate of contact between residents relative to the rate of contact involving staff members (which is not affected by general isolation). In our main results, we assume that these proportions are the same (with each rate set to an average of three per resident per day), which is consistent with empirical estimates for residential care facilities (see Discussion). However, our sensitivity analysis demonstrates that if resident-resident contact rates increase, our model produces a corresponding increase in the marginal impact of general isolation measures with lower relative increases observed for lower reproductive ratios (see Supporting Information, Figure~\ref{fig:bkgCR_SA}). 

Another potential sensitivity of our model arises from the assumption that compliance with targeted case isolation is fixed at 90\%. In our sensitivity analysis (see Supporting Information Figure \ref{fig:CIC_SA}), we demonstrate that our results about the marginal benefits of general isolation are robust to this assumption, as long as the condition is satisfied that general isolation compliance cannot be greater than compliance with targeted case isolation. While outbreak sizes increase when targeted isolation compliance decreases, the above condition prevents general isolation from providing a substantial additional benefit. 
}

To supplement the findings detailed above, our full results are summarised below in {Table~\ref{tab:medians}}.
\setlength{\tabcolsep}{5pt}
\begin{table}
    \centering
    \begin{tabular}{p{0.2\linewidth}|p{0.1\linewidth}|p{0.1\linewidth}|p{0.1\linewidth}|p{0.1\linewidth}|p{0.1\linewidth}}
    screening strategy&general isolation level&cum. infections&outbreak duration&peak FTE deficit&cum. residents isolated\\
    \hline
    full asymptomatic screening&  90\%&  14 (11)&  11 (8)&  3.4 (6.2)&  6 (4)\\
    \hline
    full asymptomatic screening&  50\%&  15 (12) &  12 (10) &  3.6 (6.2)&  6 (5)\\
    \hline
    full asymptomatic screening&  0\%&  17 (14.5)&  14 (13)&  3.6 (6)&  7 (8)\\
    \hline
    asymptomatic screening during outbreaks&  90\%&  29 (26)&  14 (7)&  6.6 (11.8)&  12 (11)\\
    \hline
    asymptomatic screening during outbreaks&  50\%&  31 (28)&  15 (10.5)&  6.4 (12)&  13 (13)\\
    \hline
    asymptomatic screening during outbreaks&  0\%&  33 (32)&  16 (14)&  6.6 (12))&  15 (17)\\
    \hline
    no asymptomatic screening&  90\%&  32.5 (31)&  16 (8)&  4.4 (6.6)&  9 (7)\\
    \hline
    no asymptomatic screening&  50\%&  37 (37)&  20 (12)&  4.2 (7.2)&  11 (12)\\
    \hline
    no asymptomatic screening& 0\%& 44.5 (53)& 24 (18)& 4.4 (6.6)& 15 (22)\\
    \hline
    unmitigated outbreaks
    &  -&  $173^*$ (20)&  -&  -&  -\\
    \hline
    \end{tabular}
    
    {{$^*$ Median over all unmitigated outbreaks which produced more than 80 infections (corresponds to the second mode of the distribution shown in Figure \ref{fig:I_tot_CIPPE_vs_UNMIT}a).}}
   
    \caption{Summary statistics for model output over a range of mitigation scenarios. Medians values are shown for cumulative infections, outbreak duration, peak FTE deficit, and the cumulative number residents placed into case isolation. {\color{purple}{Next to each median value, the interquartile range is shown in parenthesis.}}}
    \label{tab:medians}
\end{table}

\FloatBarrier

\section{Discussion}

Residential aged care environments are uniquely challenging settings for infection control and prevention \cite{usher2021preparedness}. The combination of high medical risk, prevalence of cognitive impairment, high service needs, and insecure workforce produces conditions in which outbreaks are more likely, infections cause more severe illness, and transmission is difficult to control. During the COVID-19 pandemic, outbreaks in these environments caused large numbers of deaths worldwide \cite{shallcross2021factors,comas2020mortality}. Simultaneously, the measures taken to mitigate outbreaks caused disproportionate physical and mental health burdens on residents \cite{dyer2022managing,sweeney2022experiences,liddell2021isolating,Haj2020high}. Trade-offs associated with resident and workforce well-being were especially pronounced before pharmaceutical interventions such as vaccines and antiviral drugs were available \cite{stratil2021non,sims2022social}. While the availability of antivirals and vaccines have substantially reduced the pressure to apply nonpharmaceutical interventions (NPIs) in aged care environments, the prospect of future pandemics or the evolution of SARS-CoV-2 variants of concern justifies a retrospective focus on the application of NPIs. 

If disease transmission modelling is to be applied effectively to inform outbreak response in such settings, models must effectively capture the semi-structured nature of the RACF environment and realistic constraints on the effectiveness of mitigation strategies. In this work, we described and demonstrated an agent-based simulator which can account for salient factors such as staff roster scheduling and non-random assignment of staff to residents, while also allowing random interactions between residents. To account for the physical, ethical, and logistical constraints arising from the uncertain behaviour of residents and from the material challenges associated with testing and PPE use, the model allows variation in compliance with scheduled testing, case isolation, and general isolation. It also allows the efficacy of PPE to vary depending on the types of individuals interacting. Another important constraint that our model accounts for is the imperfect and time-varying sensitivity of tests, which limits the degree to which isolation of detected cases can mitigate presymptomatic transmission. For realistic case detection probabilities, we implemented a previously-validated sub-model of test sensitivity as a function of time-since-infection to capture the sensitivity of tests occurring early during an infection (i.e., before expression of symptoms) \cite{ZachresonQuar2022}. With the features described above, the model was well suited to our investigation of general isolation policies and asymptomatic testing rates. 

While general isolation measures may help mitigate transmission, the mental health toll produced by such measures, especially to residents near the end of life or with cognitive impairment, calls into question whether their application is beneficent. For them to be justified, it must be demonstrated that such measures provide substantial mitigation of transmission supplemental to what is achieved by those measures which may be applied at lower cost to general well-being. From this perspective, our results do not support the implementation of general isolation conditions to individuals without confirmed infection. Rather, we show that the benefits of combining (imperfect) case isolation with the deployment of PPE upon declaration of an outbreak render negligible the additional benefit achieved through general isolation measures. 

Few previous modelling studies examine general isolation policies. Most recent studies focus on the role of vaccination and infection surveillance \cite{gomez2022testing,smith2020optimizing,lasser2021agent,litwin2022preventing}, with some examining case isolation \cite{rosello2022impact}, cohorting \cite{nguyen2021impact}, and furlough of staff \cite{fosdick2022model}). However, our results contrast with one other study of which we are aware. The results reported by Love {\it{et al.}} show substantial benefits from reductions in contact rates between residents, even in combination with case isolation and other measures \cite{love2021continued}. However, their model assumes very large differences in the level of contact between residents with and without social distancing (respectively: 2 vs. 50 potentially infectious contacts per resident, per day). The unmitigated contact rate (50 per day) is much higher than those reported by Vilches {\it{et al.}} (6.8 contacts per day) for Canadian care facilities \cite{vilches2021multifaceted}, and Smith {\it{et al.}} (five per day) for a French facility \cite{smith2020optimizing}, both of which are closer to the rate assumed in our implementation (three per day). Therefore, while their conclusions may be valid for scenarios with very high baseline contact rates, these rates are not realistic for residential aged care facilities.  

Our conclusions differ also from those presented by Smith {\it{et al.}}, which focused on optimal infection surveillance under resource-limited conditions \cite{smith2020optimizing}. They conclude that their results emphasize ``the importance of interventions to limit contact between patients (e.g. social distancing among retirement home residents...'', an interpretation which conflicts with our findings. However, the study by Smith {\it{et al.}} did not explicitly examine the effectiveness of distancing policies, so while their cautionary conclusions may be qualitatively consistent with their observations of model dynamics, these conclusions have no direct relationship to their quantitative findings. While we agree that general isolation protocols may be justified in the absence of sufficient surveillance capacity, our work illustrates that case isolation achieved through active screening may be sufficient to avoid the imposition of these potentially more harmful policies. 

Asymptomatic screening is a less controversial practice than general isolation measures, but still carries many caveats. While not as acutely damaging as social isolation, the continuous imposition of physically invasive tests on staff and residents who show no signs of illness may be unjustified, especially during periods of lower community prevalence \cite{kain2021routine,brust2023asymptomatic,mahase2022covid}. Furthermore, frequent asymptomatic testing may not be feasible due to resource constraints \cite{dumyati2020does}.

Here, we compared screening strategies employing differing levels of asymptomatic testing, and demonstrate marked improvements for scenarios in which asymptomatic screening is employed prior to declaration of an outbreak. These results are in general agreement with reports from real-world outbreaks (e.g., \cite{telford2020preventing}). This result comes with the caveat that our simulations are initialised with the importation of an infectious case into the facility (infecting either a staff member or a resident). Therefore, while effective at preventing outbreaks from occurring and at staunching them earlier if they do occur, our findings are consistent with the choice to implement continuous asymptomatic screening only when community transmission presents a substantial risk of case importation. We note that while the introduction of asymptomatic screening upon outbreak declaration can decrease the final size of outbreaks, this policy naturally leads to more staff absenteeism (Figure \ref{fig:CIPPE_vs_OB_only_vs_Asymp}b). {\color{purple} }

\subsection{Limitations}
  
While our model is realistic in {\color{purple} its representation of key factors such as the facility structure, staffing schedules, within-host disease dynamics, test sensitivity, and outbreak response}, in other ways the level of abstraction it applies {\color{purple} introduces some limitations}. One {\color{purple}of these is} the way in which the model simulates furlough of staff members with confirmed infection. The mechanism of case isolation for staff members is to remove them from the facility workforce for a 7-day period. While this response and its duration are realistic, we do not simulate re-assignment of rooms to the remaining staff members, an omission which could produce unrealistic reductions in network density and cause the model to over-estimate the effectiveness of furloughing staff with confirmed infection. To realistically capture labour re-distribution patterns (i.e., ``surge rostering'') additional consultation is required because specific industry standards are not available. {\color{purple} This limitation allows residents to be completely isolated due to the furlough of all staff servicing their room. We confirmed that such scenarios arise to a negligible extent (and do not occur at all in more than 90\% of simulations, see Supporting Information Figure \ref{fig:net_iso_hist}).}  

Regarding the social interface of the RACF with the outside community, our model simulates only a single importation event. We do not explicitly simulate visitors to the facility, or the transmission patterns associated with staff members outside of the facility. {\color{purple} Our model therefore implicitly assumes that visitors play a negligible role in transmission during active outbreaks, and that there is a low likelihood of multiple importations occurring prior to outbreak declaration. In scenarios with high community prevalence, multiple importation events from visitors could feasibly occur and could lead to larger numbers of presymptomatic or asymptomatic infections at the time of outbreak declaration. However, multiple importations would not alter the reproductive ratio, so we do not expect our main results regarding the marginal impact of general isolation after outbreak detection to be sensitive to the number of initial infections. Looking to future work, we note that restrictions on visitation were among the most heavily criticised policies applied to aged care facilities during COVID-19 and this subject deserves further attention focused on balancing the well-being of residents and their families with the need to prevent importation of infectious diseases.} In general, while simulating the community external to the facility is beyond the scope of this study, it should be considered for future work on topics such as optimising the criteria for outbreak declaration so to avoid false alarms, or the risks produced by RACF staff working across multiple facilities. Though these topics have been studied by others (e.g., \cite{nguyen2022hybrid}), understanding the role of staffing practices in outbreak prevention and mitigation is applicable to other healthcare sectors as well and many challenges remain unaddressed \cite{willan2020challenges,jeleff2022occupational}.

Incorporating the effects of vaccines and prior exposure on transmission potential is feasible with the model described here but was not implemented in this study, in which we placed the focus instead on non-pharmaceutical interventions. We made this choice in part because of the great diversity of exposure and vaccination histories that currently exist in aged care facilities, and in part because we aimed to present results with generic implications beyond COVID-19. 

Finally, our choice to assume a uniform asymptomatic fraction of 33\% neglects the potential for SARS-CoV-2 infection to present higher clinical fractions in older age groups. Assuming higher symptomatic proportions in the resident population due to their older age could reduce the marginal impact of asymptomatic screening. However, even with a low asymptomatic fraction, screening would still play an important role in early case identification during incubation, limiting the sensitivity of our findings to this simplifying assumption.

\section{Conclusion}

In conclusion, our study used an agent-based simulation model to examine realistic combinations of policies for COVID-19 outbreak mitigation within residential aged care facilities. We focused our analysis on assessment of asymptomatic screening strategies, and general isolation policies for outbreak response. Our findings are consistent with the position that general isolation of residents is not justified if screening resources are sufficient for frequent surveillance, PPE is available to staff and residents, and case isolation can be conducted according to clinical practice guidelines. Furthermore, our results support the recommendation for continuous asymptomatic screening of residents and staff, with the caveats that such measures will be subject to resource constraints and analysis of case importation risk. Future work extending this study should focus on the role of hybrid immunity from vaccination and virus exposure, the facility-community interface, and the design of industry workforce practices that can help limit the likelihood of case importation. 

\section{List of Abbreviations}
RACF - residential aged care facility; PPE - personal protective equipment; NPIs - nonpharmaceutical interventions; FTE - full-time equivalent; 

\section{Declarations}

\subsection{Ethics approval and consent to participate}
Not applicable

\subsection{Consent for publication}
Not applicable

\subsection{Availability of data and materials}
All code required for reproduction of the results reported in this study is contained in the associated GitHub repository \url{https://github.com/cjzachreson/RACF_C19_A}. 

\subsection{Competing Interests}
Cameron Zachreson: Reports no competing interests. \\

Ruarai Tobin: Reports no competing interests.\\

Camelia Walker: Reports no competing interests. \\

Eamon Conway: Reports no competing interests.\\ 

Freya M Shearer: Reports payments made to the University of Melbourne by DMTC Limited for supporting the development of an epidemic decision support system. \\

Jodie McVernon: Reports payments to the University of Melbourne by the World Health Organisation and payments to the University of Melbourne by the Australian Department of Foreign Affairs and Trade. \\

Nicholas Geard: Reports no competing interests\\

\subsection{Funding}
This work was funded by the Australian Department of Health and Aged Care {\color{purple} and by the Australian Government’s Medical Research Future Fund (MRF2017355).}

\subsection{Author Contributions}
CZ composed the original manuscript and figures, composed all simulation code, performed the numerical experiments and composed the figures. All authors contributed to model development. JMcV and NG conceived of the study and supervised the research. All authors contributed to manuscript drafting and editing.  

\subsection{Acknowledgements}
We would like to acknowledge Jacob Madden for facilitating consultation with RACF representatives and professionals.

\section{References}
\bibliography{references.bib}

\FloatBarrier
\clearpage
\newpage

\newcommand{\beginsupplement}{%

 \setcounter{table}{0}
   \renewcommand{\thetable}{S\arabic{table}}%
   
     \setcounter{figure}{0}
      \renewcommand{\thefigure}{S\arabic{figure}}%

       \setcounter{algorithm}{0}
      \renewcommand{\thealgorithm}{S\arabic{algorithm}}%
      
      \setcounter{page}{1}
      \renewcommand{\thepage}{S\arabic{page}} 
      
      \setcounter{section}{0}
      \renewcommand{\thesection}{S\arabic{section}}
      
      \setcounter{equation}{0}
      \renewcommand{\theequation}{S\arabic{equation}}
     }

\beginsupplement

\FloatBarrier
{\bf \Large{Supporting Information}}

\section{Detailed methods}

{\color{blue}
\subsection{Overview}

The model used in this work consists of three main components: 1) a structured population contact network model based on the characteristics of an RACF environment, 2) a model of COVID-19 within-host disease progression and transmission between infected and susceptible individuals, and 3) a model of case detection and outbreak response. Pesudocode describing the overall simulation algorithm is shown in Algorithm \ref{alg:overview} and a more detailed description of the transmission algorithm providing key details of the transmission model implementation is provided in Algorithm \ref{alg:transmission_simulation}. 

A list of key model parameters is provided in Tables \ref{tab:params_pop_model}, \ref{tab:params_c19_model}, and \ref{tab:params_OB_response}. The method descriptions provided in the following sections provide details of model implementation.} {\color{purple} The method description below does not contain mathematical details of the dynamic model of test sensitivity, which is described comprehensively in our previous work \cite{ZachresonQuar2022}.}

\setlength{\arrayrulewidth}{0.25mm}

\begin{figure}
\begin{algorithm}[H]
\caption{Overview of the model algorithm.}\label{alg:overview}
\begin{algorithmic}
\State population $\gets$ \Call{generate population}{facility characteristics}
\State network $\gets$ \Call{initialise contact network}{population}
\State scenario $\gets$ \Call{set scenario configuration}{input parameters}
\State configuration $\gets$ (population, network, scenario)
\While{outbreaks $<$ 1000}
    \State initial conditions $\gets$ \Call{initialise dynamics}{configuration}
    \State (outbreak declared, stats) $\gets$ \Call{transmission}{initial conditions, configuration}
    \If{outbreak declared}
        \State{outbreaks $\gets$ outbreaks + 1}
        \State{push stats to ensemble output}
    \EndIf
\EndWhile
\State \Return summary statistics for all outbreaks
\end{algorithmic}
\end{algorithm}
\end{figure}

\begin{figure}
\begin{algorithm}[H]
\caption{Outbreak simulation algorithm}\label{alg:transmission_simulation}
\begin{algorithmic}
    \Function{transmission}{initial conditions, configuration}
        \State outbreak declared $\gets$ false
        \State termination flag $\gets$ false
        \While{termination flag is false}
            \State {\bf{iterate transmission dynamics:}}  
            \State $t \gets t + \Delta t$
            \State \Call{update active infections}{population} \Comment progress within-host model
           
            \\\hrulefill
       
            \State {\bf{update environment:}}
            \If{new day}
                \State \Call {update structured contact network}{day of week, network}
                \State \Call{test staff and residents}{outbreak declared, scenario, population}
                \State \Call{update case isolation}{outbreak declared, scenario, population}
                
                \\\hrulefill
             
                \State {\bf{outbreak response:}}
                \If{outbreak declaration conditions are met}
                    \State {outbreak declared $\gets$ true}
                    \State \Call{outbreak response}{scenario} \Comment{alters testing, PPE, distancing}
                \EndIf
                \If{termination conditions are met}
                    \State{termination flag $\gets$ true}
                \EndIf
            \EndIf

           \\ \hrulefill
          
            \State {\bf{contact sampling:}} 
            \State initialise potential transmission pairs: $E_I \gets [~]$
            \State initialise aggregate weight of infectious edges: $w_I \gets 0$
            \For{$a$ $\in$ infected agents}
                \State push edges $e_a$ to $E_I$ \Comment{compile list of infectious edges}
                \State $w_I \gets w_I + \sum\limits_{e \in e_a} w_e$ \Comment{sum weights of infectious edges}
            \EndFor
            \State sample infectious edges: $E_T \subset E_I : p(e \in E_T) \propto w_e/w_I $
            \State push background contacts to $E_T$ \Comment{random contacts between residents}
          
          \\\hrulefill
         
            \State {\bf{pariwise transmission:}}
            \For{$e \in E_T$}
                \State evaluate pairwise transmission probability
                \If {transmission successful} infect susceptible contact \EndIf
            \EndFor
        \EndWhile 
        \State \Return{outbreak declared, summary stats}
    \EndFunction
\end{algorithmic}
\end{algorithm}
\end{figure}

\FloatBarrier


\begin{table}
    \centering
    \begin{tabular}{p{0.1\linewidth} p{0.6\linewidth} p{0.15\linewidth} p{0.1\linewidth}}
         \hline
         \multicolumn{3}{l}{Parameters: RACF facility model}\\
         \hline
         Symbol &Description &Value \\
         \hline
         $n$ &Number of residents & 88\\
         $k$ & Number of facility staff & 121\\
         $p_{med}$ & Prop. of medically-trained staff & 0.3\\
         $p_{needs}$ & Prop. of high-needs residents & 0 \\
         $p_{shared}$ & Prop. of double-occupancy rooms & 0 \\
         $p_{\text{min}}$ & \makecell[l]{minimum fraction of staff present \\ each day in baseline roster} & 0.2 \\
         $p_{5}$ & \makecell[l]{prop. of staff working 5 days per week}  & 0.26 \\
         $p_{3}$ & \makecell[l]{prop. of staff working 3 days per week} & 0.36 \\
         $p_{2}$ & \makecell[l]{prop. of staff working 2 days per week} & 0.38 \\ 
         $k_{gen}$ & \makecell[l]{number of general staff \\ assigned to each room per day} & 4 \\[2ex]
         $k_{med}$ & \makecell[l]{number of medical staff \\ assigned to each room per day} & 1 \\ 
         \hline
         \multicolumn{3}{l}{Structured contact network}\\
         \hline
         $c$ & mean number of contacts per resident, per day & 3.0\\
         $\lambda_{\text{(i)}}$ & mean total contact rate (contacts per day) & $nc$\\
         $w_e$ & sampling weight for each edge & see Table \ref{tab:network_weights} \\
         \hline
         \multicolumn{3}{l}{Unstructured contact network}\\
         \hline
         $\lambda_{\text{(ii)}}$ & \makecell[l]{mean number of background contacts \\ per resident per day (baseline, no active outbreak)} & 3.0 \\
         \hline
   
    \end{tabular}
    \caption{Summary of key model parameters related to population structure. {\color{purple}For additional details regarding the facility model and contact networks see the corresponding sections below.}}.\label{tab:params_pop_model}
\end{table}

\begin{table}
    \centering
    \begin{tabular}{p{0.1\linewidth} p{0.5\linewidth} p{0.3\linewidth}}
         \hline
         \multicolumn{3}{l}{Parameters: COVID-19 transmission model}\\
         \hline
         Symbol &Description &Value \\
         \hline
         $\kappa$ & global transmission scalar & 0.2 \\
         $R_0$ & mean reproductive ratio & $\approx 2.4$ \\
         $\delta$ & \makecell[l]{dispersion parameter of  \\ secondary case distribution} & 0.1 \\
         $\beta_{max}$ & \makecell[l]{random variable determining \\ an individual's peak infectiousness} & \makecell[l]{Gamma\\(shape = $\delta$, scale = $\kappa / \delta$)}\\
         $T_{\text{inc}}$ & mean incubation period & 5.5 days\\
         $t_{\text{inc}}$ & \makecell[l]{random variable determining \\ an individual's incubation period} & \makecell[l]{Lognormal \\ ($\mu = 1.62$, $\sigma = 0.418$)}\\
         $T_{\text{rec}}$ & mean recovery period & 7.5 days  \\
         $t_{\text{rec}}$ & an individual's recovery period & \makecell[l]{Uniform(5, 10)}\\
         $p_{\text{asymp}}$ & \makecell[l]{mean proportion of\\ asymptomatic infections} & 0.33 \\

         \hline
    \end{tabular}
    \caption{Summary of key model parameters. {\color{purple}For additional details on the COVID-19 transmission model, see the relevant section below and our previous work \cite{ZachresonQuar2022}}.}\label{tab:params_c19_model}
\end{table}

\begin{table}
    \centering
    \begin{tabular}{p{0.1\linewidth} p{0.5\linewidth} p{0.2\linewidth} p{0.1\linewidth}}
         \hline
         \multicolumn{3}{l}{Parameters: Outbreak detection and response}\\
         \hline
         Symbol &Description &Value \\
         \hline
         $\rho_{max}$ & \makecell[l]{peak test sensitivity\\(see \cite{ZachresonQuar2022} and \cite{hellewell2021estimating} for further details)} & $\approx 0.83$ \\[2ex] 
         $\Delta^{\text{staff}}_{\text{min}}$ & \makecell[l]{minimum time (in days) \\ between asymptomatic tests \\ for staff members (no active outbreak)} & 3 \\[2ex] 
         $p^{\text{resident}}_{\text{test}}$ & \makecell[l]{probability per day of an asymptomatic \\ resident testing for infection} & \makecell[l]{0 or 1 \\ (depends on scenario)}\\
         $p^{\text{symptoms}}_{\text{test}}$ & \makecell[l]{probability of testing if symptomatic \\ (note: staff members are only tested if present)} & 1.0 \\
         $t_{\text{furlough}}$ & duration of furlough period for staff who test positive & 7 days \\
         $t_{\text{iso}}$ & duration of case isolation period for residents who test positive & 7 days \\
         $p_{\text{iso}}$ & case isolation efficacy & 0.9 \\
         $L$ & general isolation efficacy & \makecell[l]{0.9, 0.5, or 0 \\ (depends on scenario)} \\
         $\eta_{ab}$ & efficacy of infection control (e.g., PPE) & \makecell[l]{0.2, 0.5, or 0.9 \\ (depends on context\\ see Table \ref{tab:PPE_efficacy_ab}) }\\
         \hline
    \end{tabular}
    \caption{Summary of key model parameters. {\color{purple}For additional details regarding the outbreak response model (including screening strategies, criteria for declaring outbreaks and criteria for declaring outbreaks to be over) see the corresponding sections below.}}.\label{tab:params_OB_response}
\end{table}

\FloatBarrier

\subsection{Facility Characteristics}
In our model, a residential aged care facility is described by the following characteristics: 
\begin{itemize}
    \item {number of residents $n$}
    \item {number of staff $k$}
    \item {proportion of staff who are medically trained $p_{med}$}
    \item {proportion of residents with high needs $p_{needs}$}
    \item {proportion of shared rooms $p_{shared}$}
\end{itemize}
For the results reported in this study, we simulated a single facility with the values for these parameters listed in Table \ref{tab:params_pop_model}. The number of rooms is determined by the number of residents and the proportion of shared rooms, and assumes that all shared rooms are double-occupancy (Note that in this work we assume all rooms are single-occupancy). 

Assignment of staff to a work roster follows the data on full-time-equivalent (FTE) and number of sector employees described in the 2020 census of Australian aged care facilities \cite{RACFcensus}. Rooms are allocated to staff members based on their role (medical or general staff) and are allocated to ensure all rooms are serviced while distributing the number of room assignments as evenly as possible among the workers available on each day of the roster period (1-week). More details are provided below: 
\begin{enumerate}
    \item {{\bf Staff types:} All staff are either Personal Care Workers (PCWs), or nurses and allied health professionals (medical staff). These all fall under the category of ``direct care staff''. In other words, we do not model the activities of ``ancillary'' staff (i.e., cooking, laundry, cleaning).}
    
    \item {{\bf Medically trained staff:} We assume that 30\% of the aged care workforce is medically trained (nurses or allied healthcare). This assumption is based on the headcount numbers in the 2020 Aged Care Workforce Census Table~2.2 \cite{RACFcensus}}
    
    \item{{\bf Minimum staffing (under normal conditions):} Under normal conditions (no active outbreak) the number of staff present at the facility may not be less than 1/5 the total headcount (i.e., if a facility has 100 PCWs, a roster will not be simulated that has fewer than 20 of them present on any given day). This constraint is applied separately to the medical staff and PCWs, and exists to ensure that rosters are not generated with unrealistically low numbers of staff present.}
    
    \item{{\bf Staff roster:} The staff roster repeats weekly, and is generated by assuming 26\% of staff work 5 days/wk, 36\% work three days per week, and 38\% work two days per week. This is derived from the headcount totals and FTE (see the 2020 Aged Care Census report Table~2.2 \cite{RACFcensus}).}
    
    \item{{\bf Assignment of rooms to staff:} On a given day, each room of a facility is serviced by exactly 4 PCWs and 1 medical staff member. The allocation of rooms to staff members attempts to produce uniform labour distribution (though the specific constraints applied to individual facilities require some relaxation of this rule). This is done to ensure that all rooms are serviced by multiple staff members, and that individual staff members are not allocated a disproportionate number of rooms.}
    
    \item{{\bf Assignment of rooms to residents:} Each resident is assigned to one room.}
    
    
    
    \item{{\bf Consistency in the assignment of staff to sets of rooms:} To the extent allowed by the intermittent work schedule defined by the roster (and the staffing requirement of each room), staff are assigned the same set of rooms on each day they are present at the facility. This is designed to simulate the tendency for staff to be assigned consistently to the same locations, and can be adjusted to produce increased levels of randomness in room assignments from day to day.}
\end{enumerate}

\subsection{within-host model of disease progression} 
The within host model of disease transmission features infectiousness that increases from the time of infection, peaking closely before symptom onset, and then declining until recovery. Crucially, we assume an over-dispersed distribution of peak infectiousness values which corresponds to a negative binomial secondary case distribution, if a sufficiently large number of susceptible contacts exist.

Each infected individual's infectiousness varies with the time since infection $\tau$. The trajectory is a 3-part piece-wise function described by an initial increase, a brief plateau (spanning the end of the incubation period), and a decline towards recovery: 
\begin{equation}\label{eq:infectivity}
\beta(\tau)  =
 \begin{cases} 
     \frac{\beta_{\text{max}}}{V_{\text{max}}} \big[\exp(k_{1}\tau) - 1\big] & \tau\leq t_{\text{inc}} - T_{p} \\
      \beta_{\text{max}} & t_{\text{inc}} - T_p \leq \tau < t_{\text{inc}}\\
      \beta_{\text{max}}\big[\exp\big(k_{2}[\tau-t_{\text{inc}}]\big) - [V_{\text{max}}]^{-1}\big] & \tau \geq t_{\text{inc}}
   \end{cases}\,,
\end{equation}
where $t_{\text{inc}}$ is the incubation period of the individual, $T_p$ is the duration of the infectiousness plateau (set equal to $0.1t_{\text{inc}}$), and $\beta_{\text{max}}$ is the maximum infectiousness of that individual. The parameter $V_{\text{max}}$ is controls the shape of the growth curve (smaller values of $V_{\text{max}}$ produce broader growth and decay functions, higher values produce steeper growth and decay). The rate parameters $k_1$ and $k_2$ are determined by the value of $V_{\text{max}}$, and the duration of incubation and post-incubation periods (which are sampled for each infected individual): 
\begin{equation}
    k_1 = \ln(V_{\text{max}})~[t_{\text{inc}} - T_p]^{-1}\,,
\end{equation}
and
\begin{equation}\label{eq:k2}
    k_2 = \ln\big([V_{\text{max}}]^{-1}\big)~t_{\text{r}}^{-1}\,,
\end{equation}
where $t_r$ is the time between symptom onset and recovery, which is drawn uniformly at random from the range [5d, 10d], to capture a plausible range for the duration of infectiousness after symptom onset {\color{blue} with a mean of $T_{\text{rec}} = 7.5$ days.}

In the individual-level model of infectiousness described above, the incubation period plays a central role in determining the dynamics of infectiousness. For each individual, these are drawn at random from a lognormal distribution with log-mean $\mu_{inc} = 1.62$ and $\sigma = 0.418$ (for a mean incubation period of $T_{\text{inc}} = 5.5$ days).  {\color{blue} For asymptomatic infections, the time-dependent infectiousness trajectory is computed identically as for those which do express symptoms. The only difference is that symptoms cannot be observed after incubation and symptomatic screening for case detection is not conducted. We simulate a mean asymptomatic fraction of $33\%$ using a probabilistic implementation: at the moment of infection, an individual will be labeled as asymptomatic with a probability of $p_{\text{asymp}} = 0.33$. {\color{purple} This asymptomatic fraction is in line with global estimates but we do not account for heterogeneity by age  \cite{shang2022percentage,ma2021global,oran2021proportion,sah2021asymptomatic}}. }

To produce the over-dispersed secondary case distribution shown in Figure \ref{fig:overdisp}, the individual-level parameter $\beta_{max}$ is drawn from a Gamma distribution with shape parameter $k_{\text{shape}} = 0.1$ and scale parameter $s = \kappa/k_{\text{shape}}$, where the mean $\kappa$ globally scales the transmission rate and reproductive ratio. {\color{blue} This implementation produces a negative-binomial secondary case distribution with dispersion $\delta = 0.1$, equivalent to the shape parameter of the Gamma distribution for $\beta_{max}$.}

Additional details of the model of within-host infectiousness, including calibration, can be found in our previous work \cite{ZachresonQuar2022}.

\subsection{calibration of the reproductive ratio}
\begin{figure}[h]
    \centering
    \includegraphics[width=\textwidth]{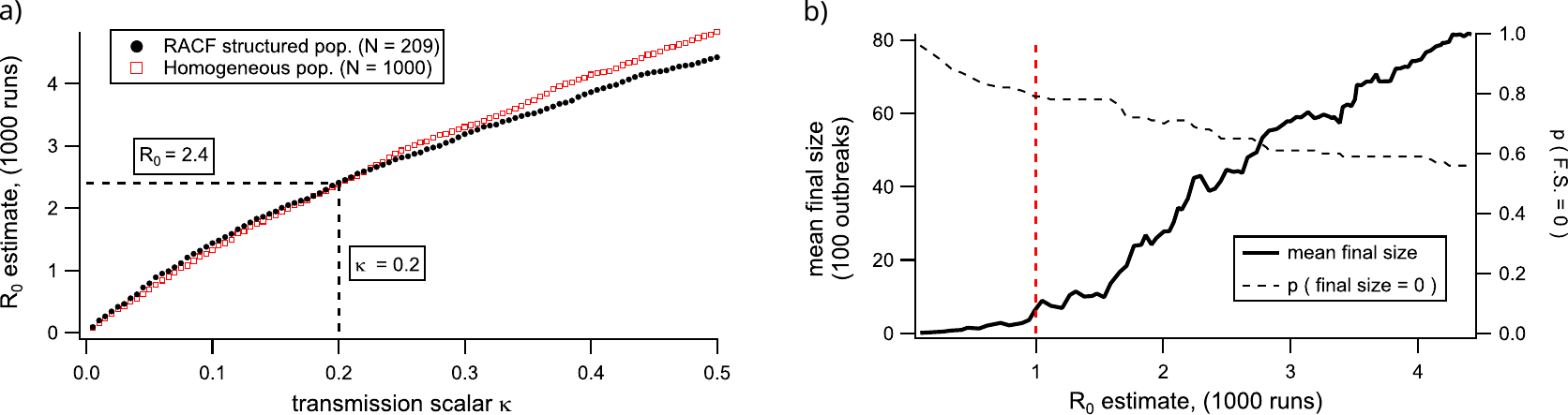}
    \caption{Reproductive ratios (a) and mean final size (b) as a function of the global transmission scalar $\kappa$. The reproductive ratios in (a) are computed over 1000 independent index case simulations for the RACF population used for the main results (black dots) and also for a homogeneous population of 1000 individuals (red squares). The $R_0$ value used in our study is shown by the dashed lines in (a). The final size statistics in (b) are computed over 100 introductions and include those which do not meet the criteria for outbreak declaration (i.e., they include those simulations which are subject to stochastic die-out). The dashed trace in (b) indicates the probability that no secondary cases will be generated, for each value of $\kappa$ [right y axis in (b)]. }
    \label{fig:R0_vs_kappa}
\end{figure}

\begin{figure}[h]
    \centering
    \includegraphics[width=0.75\textwidth]{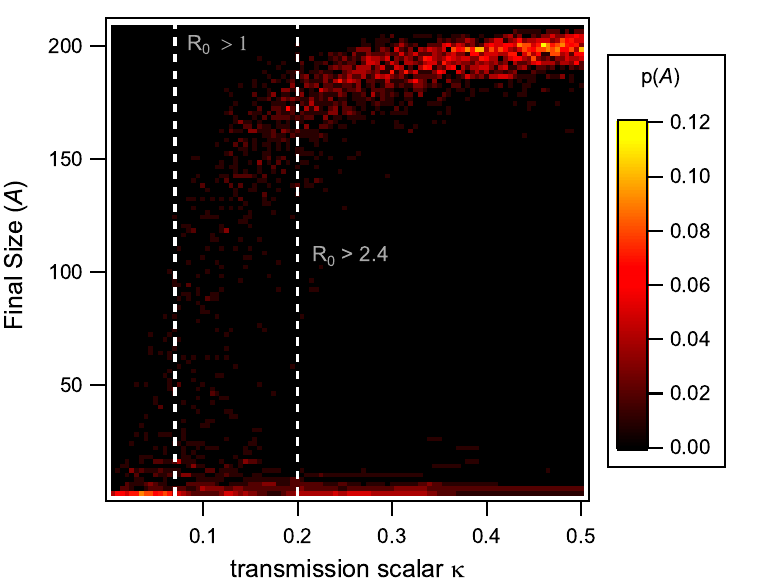}
    \caption{Heat map of final size probabilities as a function of the global transmission scalar $\kappa$. Because we employ an overdispersed secondary case distribution, this heatmap excludes $p(A = 0)$ to facilitate visualisation (note that $p(A = 0)$ is shown in Figure \ref{fig:R0_vs_kappa}b ). Dashed lines correspond to the theoretical critical threshold $R_0 = 1$ and the reproductive ratio used for our main results $R_0 \approx 2.4$. }
    \label{fig:FS_dist_vs_kappa}
\end{figure}

To calibrate the basic reproductive ratio, we systematically varied the global transmission scalar $\kappa$. For each value of $\kappa$, we simulated an ensemble of 1000 index cases and recorded the number of secondary cases produced for each instance. The mean of each ensemble provided an estimate of the reproductive ratio for the corresponding value of $\kappa$. Figures describing the calibration of $R_0(\kappa)$ as well as final size statistics for reach value of $\kappa$ are shown in Figure \ref{fig:R0_vs_kappa} and Figure \ref{fig:FS_dist_vs_kappa}. Based on this calibration we selected $\kappa = 0.2$ to produce a reproductive ratio $R_0 \approx 2.4$. To ensure index cases were selected commensurate with their relative likelihood of being infected during an outbreak, index case selection was weighted based on the sum of all contact edge weights associated with each agent (see below for more details about the contact network specifications).

\subsection{test sensitivity}
The time-dependent test sensitivity function is derived from a study that performed prospective serial testing of healthcare workers \cite{hellewell2021estimating} and follows a similar trajectory to infectiousness, with sensitivity peaking just before symptom onset and a gradual decline afterwards. Test sensitivity of rapid antigen tests is assumed to peak at approx. 83\% on average (though this varies from person to person). Extensive details about the specifics of the test sensitivity model used here can be found in our previous work \cite{ZachresonQuar2022}. We assume tests are 100\% specific (no false positives). 


\subsection{network model of disease transmission}
To simulate disease transmission within the facility population, we employ a contact network model. This model has two components (i) a structured network built to represent potential co-location dynamics within the rooms of the facility and (ii) a homogeneous network of random interactions between residents representing unstructured contact due to (e.g.) group activities, shared meals, or casual social interactions. 

\subsubsection{component (i): structured contact network}
Component (i) of the transmission network is a dynamic weighted multigraph (multiple edges may exist between two nodes) that includes both staff members and residents as nodes. The edges in the network connect staff members who are assigned to service the same room on the same day, residents to their roommates, and residents to the staff members who service their rooms. Figure \ref{fig:network_schematic} provides a schematic description of how this part of the network is generated, based on the assignments of residents and staff to rooms. The network topology of component (i) is dynamic, changing for each day of the simulation based on which staff are assigned to which rooms. Furlough of staff members who test positive also changes the network topology by removing all connections with furloughed staff members. These connections are reinstated after the 7-day furlough period. The edges in component (i) are weighted, to account for the relative likelihood that a given edge is sampled during the transmission simulations (described below). Specifically, edge weights depend pairwise on the types of agents they connect as shown in Table \ref{tab:network_weights}. For staff members who are assigned more than one of the same rooms, an edge exists between them for each room they share.  

\begin{table}
   \begin{tabular}{c c}
        {\underline{type of interaction}} & {\underline{weight}}\\
        {worker $\rightarrow$ worker} & {1.0 (baseline)}\\
        {resident $\rightarrow$ resident (different room)} & {1.0 (baseline)}\\
        {worker $\rightarrow$ resident (typical needs level)} & {2.0}\\
        {worker $\rightarrow$ resident (high needs)} & {6.0}\\
        {resident $\rightarrow$ resident (same room)} & {10.0}
    \end{tabular}
    \caption{Sampling weights used for the edges of the structured contact network component of the transmission model.}\label{tab:network_weights}
\end{table}

\FloatBarrier
\subsubsection{component (ii): random contacts between residents}
Component (ii) of the transmission network is implemented implicitly as a weighted mixing network. Residents contact one another at a constant rate, implemented as a Poisson-distributed number of contacts over discrete time intervals. Weights are introduced for non-uniform edge sampling in order to account for the reduced likelihood of contacting another resident who is subject to general isolation or case isolation (see below). 

\subsubsection{contact sampling}
In each discrete simulation timestep (of duration $\Delta t = 0.1$ day), pairwise transmission dynamics are evaluated between infected individuals, and sampled subsets of their potential contacts. This sampling process is split into components (i) and (ii) as described above. Here, we describe the details of our contact sampling algorithm for components (i) and (ii). We note that the network topology of component (i) is a function both of the day of the staff roster cycle and of the set of staff who are furloughed and not present at the facility. In the description below, we omit notation associated with these two aspects of network structure dynamics, in order to clearly denote the process used to sample infectious contacts from the set allowed by network structure at any time $t$. 

For component (i), sampling depends on the total contact rate for component (i) $\lambda_{\text{(i)}}$, the aggregate weight of edges in component (i) $w_{\text{(i)}}$, the set of edges connected to infected individuals $\{E_{I}\}$, and their aggregate weight $w_{I}$. The mean number of potentially infectious edges to evaluate in a timestep is then given as: 
\begin{equation}
\lambda_I = \lambda_{\text{(i)}} \frac{w_I}{w_{\text{(i)}}}\Delta t\,, \label{eq:lambda_I}
\end{equation}
i.e., the expected total number of contacts during the timestep multiplied by the proportion of infected edges. The number of potentially infectious contacts from component (i) $c_{\text{(i)}}(t)$ sampled for a given timestep is then drawn from a Poisson distribution: 
\begin{equation}
c_{\text{(i)}}(t) \sim \text{Poisson}(\lambda_I)\,,
\end{equation}
and the $c_{\text{(i)}}(t)$ contacts on which transmission is evaluated are selected at random (with replacement) from the set $\{E_I\}$ with sample probability of each edge $e \in \{E_I\}$ equal to $w_e / w_I$ where $w_e$ is the weight of edge $e$. This produces a set of edges sampled from component (i), $\{E_\text{(i)}\}$. 

For component (ii), the unstructured contacts between residents, sampling is implemented by iterating through infected residents and sampling contacts from the set of other residents in the facility population. The sampling algorithm depends on the baseline rate of casual contact {\it{per resident}} for component (ii) $\lambda_\text{(ii)}$, the general isolation level $L$, the effectiveness of case isolation $p_{\text{iso}}$, the set of residents $\{r\}$ who are infected at time $t$, and their isolation status $\{\theta\}$ (which takes a value of $\theta_i = 1$ if resident $r_i$ is isolated). For each infected resident $r_i$, the mean number of contacts sampled is given as:  
\begin{equation}
\lambda(r_i) = \lambda_\text{(ii)} (1 - \alpha_i) \Delta t\,, \label{eq:labda_ri}
\end{equation}
where 
\begin{equation}
\alpha_i = p_{\text{iso}}\theta_i + L(1-\theta_i)\,,
\end{equation}
to account for the reduction in contact produced by case isolation or general isolation of resident $r_i$ (note that case isolation effects supersede those of general isolation, but the two effects do not multiply). The number of potentially infectious contacts from component (ii) contributed by resident $r_i$, $c_{\text{(ii)}}(r_i)$, is sampled from a Poisson distribution:
\begin{equation}
c_\text{(ii)}(r_i) \sim \text{Poisson}(\lambda(r_i))\,.
\end{equation}
These $c_\text{(ii)}(r_i)$ edges are selected at random from the set of all other residents $\{r_{k\neq i}\}$ with sample probability equal to:
\begin{equation}
\frac{(1 - \alpha_k)}{\sum_{r_i \in \{r\}} (1 - \alpha_i)}\,,
\end{equation}
which, compiled over all infected residents produces the set of potentially infectious edges contributed by component (ii):  $E_{\text{(ii)}}$. The union of infectious edges from components (i) and (ii) gives the full set of edges to evaluate for disease transmission at time $t$: $\{E_T\} = \{E_{\text{(i)}}\} \cup \{E_{\text{(ii)}}\}$.

\subsection{pairwise disease transmission}
Once the set of potentially infectious contacts $\{E_T\}$ is determined through sampling, the probability of infection between infected individual $a$ and individual $b$ (who may or may not be infected) is computed for each edge as as: 
\begin{equation}\label{eq:p_trans}
p_{\text{trans}}(a, b, \tau) = [1 - \exp(-\beta_a(\tau)\eta_{ab})]S_b\,, 
\end{equation}
where $\beta_a$ is the force of infection produced by the infected agent $a$, $\tau$ is the time since individual $a$ was initially infected, $\eta_{ab}$ is a reduction factor for PPE effects (which depend on whether individuals $a$ and $b$ are residents or staff members), and $S_b$ takes a value of $0$ if individual $b$ is infected or recovered and $1$ if they are susceptible. Note that the role of the timestep $\Delta t$ is not included in Equation \ref{eq:p_trans} because it is accounted for in the contact rates described in Equations \ref{eq:lambda_I} and \ref{eq:labda_ri}. The transmission rate $\beta_a$ changes as a function of time since infection, as described by our within-host model of disease transmission which accounts for dispersion of infectiousness among infected individuals, and the timing of transmission as described in detail in our previous work \cite{ZachresonQuar2022}.

\subsection{implementation of interventions and PPE deployment}
The assumptions listed below with respect to screening for cases and response to outbreaks were informed by the advice detailed in the Australian Government Department of Health document entitled ``COVID-19 Outbreaks in Residential Care Facilities, Communicable Disease Network Australia, National Guidelines for the Prevention, Control and Public Health Management of COVID-19 Outbreaks in Residential Care Facilities'' dated February 15th, 2022 \cite{outbreak_guidelines}. [Note: the Australian Government recommendations have been modified since this study was conducted, the version of the guidelines we used to inform our model is included as Supporting Material to ensure continued access.]  

\begin{figure}[h]
    \centering
    \includegraphics[width=0.5\textwidth]{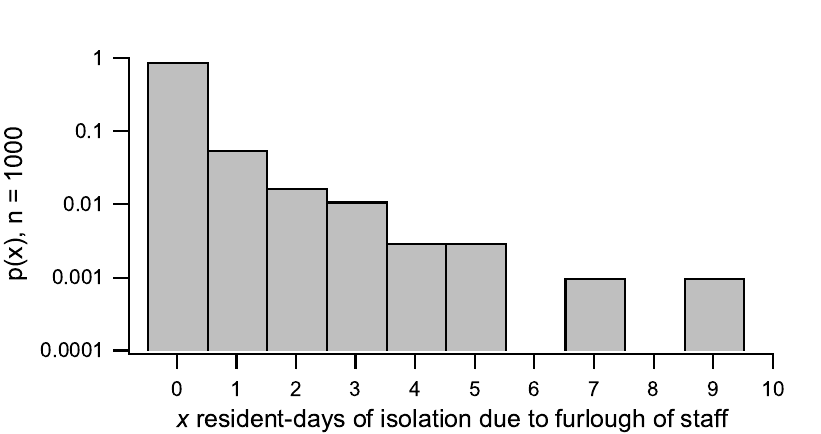}
    \caption{{\color{purple}Frequency of resident-days of isolation due to furlough of staff members over 1000 outbreaks. For these simulations, the condition producing the maximum FTE deficit was selected (asymptomatic screening during outbreaks, with 0\% general isolation level, see Table \ref{tab:medians}). Over the ensemble of 1000 simulated outbreaks, 90.1\% produced zero, 5.5\% produced one, 3.5\% produced two to five, and 0.2\% produced six to nine resident-days of isolation due to staff furlough.}}
    \label{fig:net_iso_hist}
\end{figure}

\begin{enumerate}
    \item{Staff screening (no active outbreak). Staff are tested periodically, with no more than two tests scheduled per roster period (weekly) and a minimum of $\Delta^{\text{staff}}_{\text{min}} = 3$ days between subsequent tests. This means full-time workers are scheduled for two tests per week and part time workers are scheduled for one test per week.}
    
    \item{Resident screening (no active outbreak). Residents are not subject to scheduled testing, but have a baseline probability $p^{\text{resident}}_{\text{test}}$ of being tested on any given day {\color{blue}which varies depending on the screening scenario.}}
    
    \item{Screening (active outbreak). During an active outbreak, staff are scheduled for tests on each day they are present at the facility. The probability of a resident being tested on a given day is adjusted as well, depending on the screening strategy implemented. }
    
    \item{Screening (symptom expression). Any individual who expresses symptoms (regardless of whether or not an outbreak has been declared) is tested if they are present at the facility. If they test negative, they are not required to isolate. They will be tested again on each day they express symptoms (i.e., a single negative test is not sufficient to avoid additional screening if the individual is symptomatic).}
    
    \item{Efficacy of infection control measures. After outbreaks are declared, infection control measures are put into place. These reduce the probability of transmission between different types of individuals according the the following efficacy assumptions: 

    \begin{table}
        \begin{tabular}{c c}
            {\underline{type of interaction}} & {\underline{reduction in force of infection ($\eta_{ab}$)}}\\
            {resident $\rightarrow$ resident} & {0.2}\\
            {worker $\rightarrow$ worker} & {0.5}\\
            {worker $\rightarrow$ resident} & {0.9}\\
        \end{tabular}\label{tab:PPE_efficacy_ab}
        \caption{Efficacy of infection control measures (e.g., PPE), which depends on the types of individuals interacting ($\eta_{ab}$).}
    \end{table}
    
    These choices reflect the assumption that residents use PPE and employ protective behaviour only on a discretionary basis. Staff, on the other hand, are assumed to utilise PPE to a greater extent (as per best-practice guidelines), and to almost always use PPE during interactions with residents. 
    }
        
    \item{Reduction in background contact frequency between residents during active outbreaks. The model assumes that background contact rates between residents fall by a fixed proportion during active outbreaks to simulate physical distancing, cancellation of group activities, limitations to use of communal areas (e.g., avoiding group meals), or in-room isolation policies.}
    
    \item{Throughout, we assume a 90\% reduction in background contact frequency for residents who are isolated after testing positive ($p_{\text{iso}} = 0.9$). This reflects the imperfect capacity to isolate a resident to their room at all times. [We note here that contact rates with staff and with roommates are unaffected by resident case isolation]. {\color{blue}Case isolation lasts for a fixed period of $t_{\text{iso}} = 7$ days. } }

    \item{Furlough of staff members with confirmed infection continues for $t_{\text{furlough}} = 7$ days after the day of detection. For the seven day period, a furloughed staff member is removed from the model (their connections are removed from the transmission network). We do not simulate re-distribution of room assignments upon furlough of staff, rather, we assume that all network connections not involving furloughed staff members remain unchanged. {\color{purple}This produces a small probability that residents will become completely isolated from all staff members servicing their rooms (an unrealistic condition). For quality control, we quantified this probability for the scenario producing the highest FTE deficit (this occurs when asymptomatic screening is implemented upon outbreak declaration, with general isolation level set to zero, see Table \ref{tab:medians}). The analysis summarised in Figure \ref{fig:net_iso_hist} demonstrates negligible levels: for a sample of 1000 simulated outbreaks, 90.1\% showed zero resident-days of isolation due to staff furlough, 5.5\% of simulated outbreaks produced one resident-day of isolation due to staff furlough, 3.5\% produced two to five resident-days, and 0.2\% produced six to nine resident-days of isolation due to staff furlough.}}
    
    \item{Declaring an active outbreak: For a facility to declare an active outbreak, one of two criteria must be met: (i) two new resident cases within 5 days or (ii) five new cases in staff within seven days as per the Australian Government guidelines \cite{outbreak_guidelines}. }
    
    \item{Declaring an active outbreak over: If no new resident cases have been detected within seven days, the outbreak is considered to be over, and infection control measures cease. }

\end{enumerate}

\subsection{vaccine-derived protection}

The effects of vaccination were not the focus of this study and are not explicitly simulated. However, the agent-based model implementation provided in the linked repository is capable of simulating vaccine-derived reductions in transmission. The vaccination model implemented there links vaccine-derived protection to an individually-assigned correlate of protection, using the logit-normal vaccine efficacy model described in our previous work \cite{zachreson2023individual}, with default parameters corresponding to the immune-derived protection imparted against infection with the ancestral variants of SARS-CoV-2, after previous infection. These parameters can be adjusted to account for the effects of vaccination against different variants.

\section{Calibration of $R_0$}
\FloatBarrier

\FloatBarrier

\section{Lockdown effects without case isolation or PPE}
\FloatBarrier
\begin{figure}[h]
    \centering
    \includegraphics{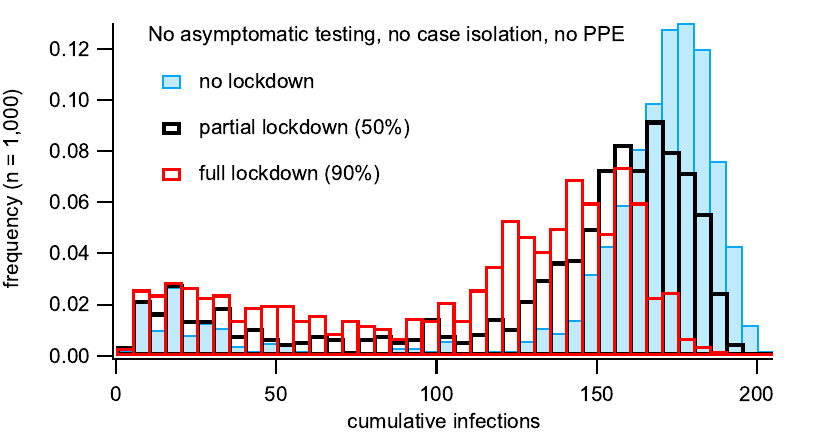}
    \caption{Effects of restricting resident-resident contact rates during outbreaks. Frequency distributions of cumulative infection numbers for each of 1000 outbreak simulations are shown for unmitigated outbreaks (blue bars), partial lockdown of residents (contact rates between residents reduced by 50\%, black bars), and full lockdown (contact rates between residents reduced by 90\%, red bars). For these simulations, all outbreak response measures other than resident lockdown were disabled (i.e., PPE, case isolation, and asymptomatic testing were not implemented). }
    \label{fig:LD_only}
\end{figure}

\FloatBarrier

{\color{purple}

\section{Sensitivity to background contact rate $\lambda_{(\text{ii})}$ and case isolation compliance}

Our main results on the marginal impact of general isolation conditions are sensitive to the rate of contact between residents (reduced through general isolation) relative to the rate of contact in the structured network (unaffected by general isolation). To investigate this sensitivity, we fix the number of structured contacts per resident fixed at $c = 3.0$, and compute outbreak final size while systematically varying the unstructured contact rate per resident ($\lambda_{(\text{ii})}$). For fair comparisons between scenarios, we fix the reproductive ratio $R_0$ by jointly varying $\kappa$ to control the transmission rate for each value of $\lambda_{(\text{ii})}$. The results in Figure~\ref{fig:bkgCR_SA} demonstrate that our primary results hold when the ratio $\lambda_{(\text{ii})}/c$ is near one. Even for larger values of $\lambda_{(\text{ii})}/c$, the marginal impact of general isolation decreases substantially as the reproductive ratio decreases (Figure~\ref{fig:bkgCR_SA}). 

The sensitivity of our main results to the level of compliance with case isolation is shown in Figure \ref{fig:CIC_SA}, which demonstrates that marginal impacts of general isolation are not sensitive to assumptions regarding case isolation compliance levels. The robustness of this result depends on the realistic assumption that general isolation compliance is bounded from above (cannot be greater than) compliance with targeted case isolation.

\begin{figure}[h]
    \centering
    \includegraphics[width = \textwidth]{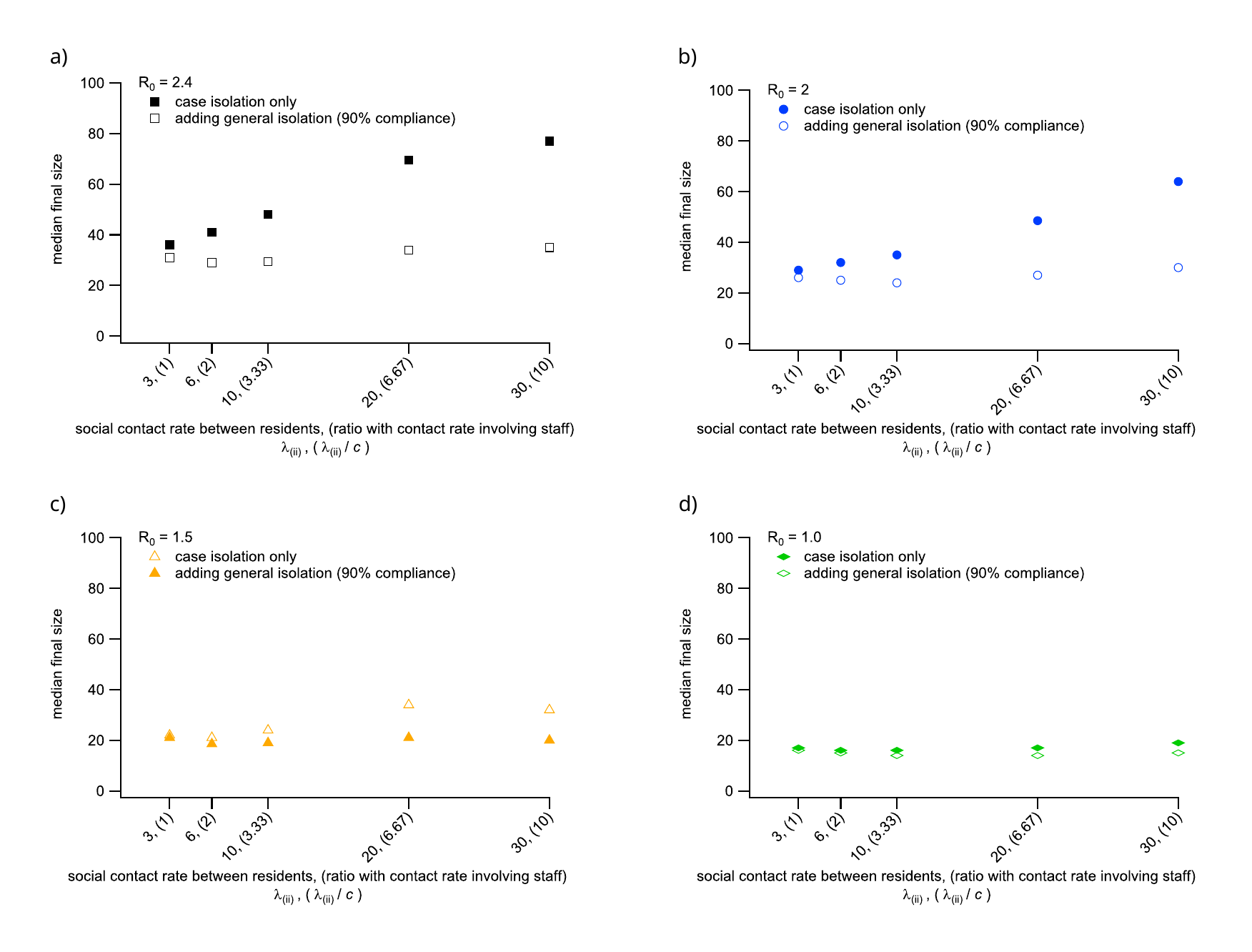}
    \caption{{\color{purple}Sensitivity analysis of background contact rate on the marginal impact of general isolation. Sample medians (n = 1000 outbreaks) are shown as functions of the ratio of the unstructured contact rate per resident $\lambda_{(\text{ii})}/c$, where $c$ is the rate of contact associated with staff members and is unaffected by general isolation measures for residents. Results are shown for different values of the reproductive ratio $R_0$ (a-d), with and without the application of general isolation for outbreak control. Marginal impacts of general isolation increase with $R_0$ and $\lambda_{(\text{ii})}$.}   }
    \label{fig:bkgCR_SA}
\end{figure}

\begin{figure}[h]
    \centering
    \includegraphics[width = 0.9\textwidth]{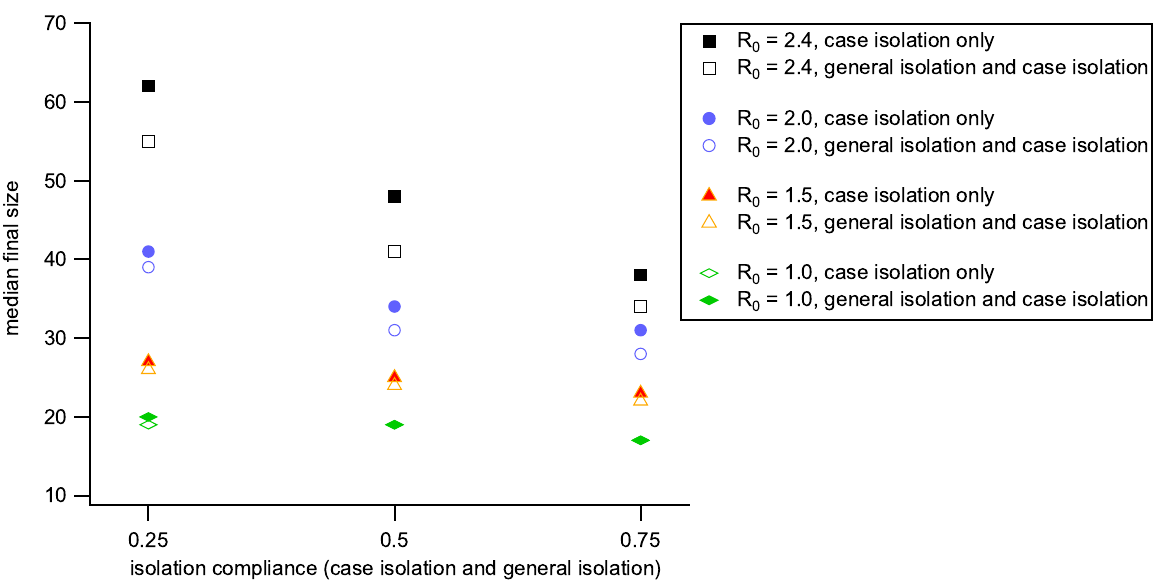}
    \caption{{\color{purple}Sensitivity analysis of case isolation compliance on the marginal impact of general isolation. Sample medians (n = 1000 outbreaks) are shown as functions of case isolation compliance rates for residents. Results are shown for four different values of $R_0$ with and without the application of general isolation for outbreak control. Here, we assume compliance with general isolation is equal to the compliance with case isolation.}}
    \label{fig:CIC_SA}
\end{figure}



}

\end{document}